\documentclass[11pt,preprint]{aastex}

\newcommand\mdot{\rm \dot{M}}
\newcommand\msun{\rm M_{\odot}}

\newcommand\lsun{\rm L_{\odot}}
\newcommand\msunyr{\rm M_{\odot}\,yr^{-1}}
\newcommand\be{\begin{equation}}
\newcommand\en{\end{equation}}

\newcommand\etal{{\rm et al}.\ }

\shorttitle{Taurus IRAC Colors}
\shortauthors{Hartmann et al.}

\begin{document}

\title{IRAC Observations of Taurus Pre-Main Sequence Stars}

\author{Lee Hartmann\altaffilmark{1}, S.T. Megeath\altaffilmark{1}, 
Lori Allen\altaffilmark{1}, Kevin Luhman\altaffilmark{1}, 
Nuria Calvet\altaffilmark{1}, Paola D'Alessio\altaffilmark{2},
Ramiro Franco-Hernandez\altaffilmark{2}, Giovanni Fazio\altaffilmark{1}}

\altaffiltext{1}{Center for Astrophysics, 60 Garden Street, Cambridge, MA 02138}
\altaffiltext{2}{Centro de Radioastronomia y Astrofisica,
Ap. P. 72-3 (Xangari) 58089 Morelia, Michoacan, Mexico}

\email{hartmann@cfa.harvard.edu}

\begin{abstract}
We present infrared photometry obtained with the IRAC camera on the {\em Spitzer} Space
Telescope of a sample of 82 pre-main sequence stars and brown dwarfs in the 
Taurus star-forming region.  We find a clear separation in some IRAC 
color-color diagrams between objects with and without disks.  
A few ``transition'' objects are noted, which 
correspond to systems in which the inner disk has been 
evacuated of small dust.  Separating pure disk systems from objects
with remnant protostellar envelopes is more difficult at IRAC wavelengths,
especially for objects with infall at low rates and large angular momenta.
Our results generally confirm the IRAC color classification scheme
used in previous papers by Allen \etal and Megeath \etal 
to distinguish between protostars, T Tauri stars with disks, 
and young stars without (inner) disks.  The observed IRAC colors are
in good agreement with recent improved disk models, and in
general accord with models for protostellar envelopes derived from
analyzing a larger wavelength region.  We also comment 
on a few Taurus objects of special interest.  Our results should be useful 
for interpreting IRAC results in other, less well-studied star-forming regions.
\end{abstract}


\section{Introduction}
The Taurus-Auriga molecular cloud has long served as a touchstone 
for understanding pre-main sequence evolution.  The proximity of Taurus,
plus the relatively low extinction in the area, has enabled researchers to develop a 
reasonably complete census of the pre-main sequence population. 
These properties also mean that the characterization of protostellar 
envelopes and circumstellar disks is more extensive and detailed 
for Taurus objects than for any other star-forming region (see, e.g.,
Kenyon \& Hartmann 1995 $=$ KH95). 

We are conducting a survey of known Taurus pre-main 
sequence stars using the IRAC camera on board the {\em Spitzer} Space Telescope.  
Although a great deal is already known about the infrared spectral energy 
distributions (SEDs) of Taurus members, the unprecedented photometric sensitivity 
of IRAC in the mid-infrared
permits the study of many more objects in the $\sim 3 - 9 \mu$m 
wavelength range than previously possible from ground-based studies. 
The IRAC data also provide an important link between measurements of
stellar photospheric emission (which generally peaks at around $1 \mu$m) and 
forthcoming detailed spectra of Taurus members obtained with the IRS 
spectrograph on {\em Spitzer} (which covers the $\sim 5-35 \mu$m range).
More broadly, our IRAC observations of previously well-characterized systems
in Taurus can serve as a guide to the interpretation of IRAC data of other,
much more distant regions which can now be studied with the 
unparalleled sensitivity of {\em Spitzer}.

In this paper we report preliminary results from our IRAC survey,
with results for 82 Taurus systems.
A fuller accounting of the $\sim 200$ pre-main 
sequence Taurus systems in the IRAC Guaranteed Time program
will be published later, as data become available.

\section{Observations}

The observations reported here were obtained with the IRAC instrument on board
the {\em Spitzer} Space Telescope (Fazio \etal 2004).
IRAC observations are made with four filters; here we refer to them by their approximate 
wavelength centers of 3.6, 4.5, 5.8, and $8 \mu$m.  A sequence of observations
in all four filter bands requires a minimum of two pointings for a given target, 
since each of the two IRAC fields of view serves serves two imaging channels.  For all of the 
observations reported here, we made three spatial dithers in each band 
to improve signal-to-noise and cosmic ray rejection.  
We used the ``High Dynamic Range'' (HDR) 12 second integration
mode, which involves taking two frames with effective exposure times of
10.4 and 0.4 seconds. 

The pre-main sequence stars in Taurus are widely dispersed on the sky.  
This means that mapping methods are inefficient
except in some regions such as L1551 and L1495 (which will be 
reported separately).  The observations reported here were mostly 
taken in ``cluster mode'', which involves identifying individual 
targets clustered within a one degree radius of a common center.
In all but a few cases, only one Taurus member (system) was observed in a single
IRAC field.  

The Taurus data were taken during two IRAC ``campaigns'' on February 9-14, 2004, and
March 6-9, 2004.  For comparison we include IRAC data for the interesting
10 Myr-old star TW Hya, taken on Dec. 20, 2003. 

The data frames were processed by the pipeline version S9.5 at the {\em Spitzer}
Science Center.
Photometry was extracted from the basic calibrated data frames using the aperture
photometry routine aper.pro in the  IDLPHOT package (Landsman 1993).  The IDLPHOT
routines were integrated into a custom IDL program which use the WCS information
in the image headers to identify the images containing a given star and find the
star within the images.  The program then obtains the centroid of the star and extracts the
photometry.   A 5 pixel aperture was used for each star, with a sky annulus
extending from 5 pixels to 10 pixels.  Since the zero points are referenced to a
star measured with a 10 pixel aperture and a sky annulus spanning radii from 10 to 20 pixels,
we scaled the photometry with
an aperture correction derived from calibration star data.  The corrections used
were  1.061, 1.064,  1.067, and 1.089
in the [3.6], [4.5], [5.8], and [8] bands, respectively.  
Instrumental magnitudes were measured from each of
the three dithered images independently.
The quoted values are the mean of the three magnitudes, and the quoted standard
deviations are the observed standard deviation in the three frames.

For binaries with separations on the order of $2-12"$, photometry was performed
by fitting point spread functions
to the two objects simultaneously using a custom IDL program.  The point spread
function (PSF) used were those
derived from the Campaign Q data, given at the {\em Spitzer} Science Center website
(http://ssc.spitzer.caltech.edu/irac/psf.html).  
For each binary system, PSF fitting was performed 
individually on all three dithered images taken.
PSF fitting photometry
was also obtained for a sample of ten,  isolated stars, and the mean offset
between the magnitudes returned by the PSF fitting and the aperture photometry
routines was measured.  The PSF fitting magnitudes were corrected by this offset
to ensure consistency with the aperture photometry reported for all the isolated
stars.   The reported uncertainty of the PSF fitting photometry is the standard 
deviation of the magnitudes determined for each of the three dithers 
added in quadrature with the uncertainty in the offset between the
aperture and PSF fitting photometry.

We adopted zero point magnitudes ($ZP$) of 19.660, 18.944, 16.880, 17.394 in the [3.6],
[4.5], [5.8] and [8] bands, respectively, where
$ [m] = -2.5 \log (DN/sec) + ZP$.   These were derived from the average calibration
of four AV star primary
standards, HD165459, NPM1p64.0581, 2MASS1812095, and NPM1p60.0581, which are observed
during every IRAC
campaign. By adopting these zero points, the  resulting IRAC colors of the A
dwarfs were equal to 0, as expected in the Vega-based IRAC magnitude system.
We estimate uncertainties in the zero point magnitudes of approximately 0.05 mag;
repeatability appears to be at the level of 0.02 mag.
                                                                                       
The IRAC data for the nearby T Tauri star TW Hya, a result of an earlier data
release, were reduced separately using a different procedure.  
Fluxes were measured in a 10 pixel radius, with an outer radius of 20 pixels 
for sky subtraction (not important for this bright object).  

\section{Results}

Table 1 lists the photometric results from IRAC, along with the dates of observation,
positions measured from the IRAC images, and 2MASS magnitudes (Cutri \etal 2003) when available.
Typical random errors in the photometry appear to be of order
0.02 - 0.03 mag, except for the faintest sources.  As mentioned in the previous section,
there may be systematic errors in the photometric zero points of $\sim$ 0.05 mag,
but because these data were obtained during two campaigns, such systematic
differences should apply relatively equally to all objects, and thus differential
magnitudes and colors among the Taurus stars should be slightly more reliable
than absolute magnitudes.

\subsection{IRAC [3.6] - [4.5] vs. [5.8] - [8] diagram and classification}

Figure 1 shows that the [3.6] - [4.5] vs. [5.8] - [8] colors of the Taurus systems
are concentrated in three fairly distinct regions.  The densest cluster of
data points reside near the origin, corresponding roughly to the nearly-zero colors expected 
for stellar photospheres in this wavelength region.  A second reasonably well-defined
set of measurements lies within [3.6] - [4.5] $\sim 0.2-0.7$, [5.8] - [8] $\sim 0.6 - 1.0$.
A third, less well-defined
(and much less populated) group of objects exhibits a similar range of [5.8] - [8] colors
but much redder [3.6] - [4.5] colors.

This grouping of points in the [3.6] - [4.5] vs. [5.8] - [8] diagram 
is related to the presence of circumstellar dust.
The principal classification scheme for pre-main sequence stars is the
Class 0-I-II-III system, which characterizes objects in terms of their infrared excesses
or spectral energy distributions (SEDs) (e.g., Lada \& Wilking 1984; Adams, Lada, \& Shu 1987;
Andre, Ward-Thompson, \& Barsony 1993, 2000).
Class 0 and I objects are thought to be protostars surrounded by dusty infalling
envelopes, and thus exhibit both relatively strong far-infrared emission and significant
near-infrared extinction from their envelopes; Class II systems are stars with disks, and
thus exhibit less infrared excess and near-infrared extinction (unless observed edge-on);
and Class III objects are essentially stars without significant amounts of circumstellar dust 
(see \S 6 for further discussion).  Based on this scheme, the 
open circles near the origin, having basically photospheric colors, should be Class III
systems; the main body of stars with significant infrared excesses should be Class II (disk)
systems; and the reddest objects in both colors should be Class 0-I systems.

We started our analysis by adopting the SED classifications given by KH95, which were
based on combining ground-based photometry with IRAS fluxes.  We revised the KH95
classifications in a few cases, for reasons described below.  In addition, for some
objects we made new classifications based on IRAC colors, to be consistent with
the overall groupings which emerged from the original KH95 classifications.

We find that the 0-I-II-III SED classes generally account for the groupings seen in Figure 1,
with a few discrepancies.  The Class III sources (open circles) lie near the origin,
with nearly zero colors, as expected; the Class II sources (filled circles) exhibit
red colors consistent with excess (dust) emission; and the reddest systems
are generally Class 0-I sources (open squares and triangles). 
The principal ambiguity occurs for Class I systems; three out of seven
such systems shown in Figure 1 lie near the red end of the distribution of Class II systems,
a result whose implications are discussed more fully in \S 6.  

In a few cases the identification of a system as Class II or Class III is also
ambiguous.  Some T Tauri stars have holes
in their inner disks; given the absence of disk dust close to the central star,
such objects may have disks emitting at far-infrared etc. wavelengths, while exhibiting
little or no near- to mid-infrared excess emission.  Thus the distinction between
``Class II'' and ``Class III'' may depend upon the wavelength of observation.

Two objects particularly illustrate this problem.
The open circle with the reddest [5.8] - [8] color ($= 0.205$) in Figure 1
represents the star CoKu Tau 4. 
Recent IRS spectra from {\em Spitzer} show
that this object has essentially no infrared excess from $\sim 5 - 8 \mu$m, while
exhibiting extremely large excess emission at wavelengths longward of $20 \mu$m. 
This spectral energy distribution
(SED) is most simply interpreted as that of a disk with a very optically thin or
evacuated inner hole, and an effective disk edge (an optically-thick disk ``wall'') 
at a radius $\sim 10$~AU (Forrest \etal 2004; D'Alessio \etal 2005a).  

The second illustrative object is TW Hya, whose position in the [3.6] - [4.5] vs. [5.8] - [8]
diagram is denoted by a solid triangle.  TW Hya has a SED that is fairly similar
to that of CoKu Tau 4, and which has been interpreted in a similar manner as a disk
system with an inner hole.  The inner radius of the optically-thick
disk wall is smaller in TW Hya than in CoKu Tau 4 ($\sim 4$~AU),  
and there evidence for a small amount of dust inside the hole (Calvet \etal 2002).
GM Aur also exhibits similar behavior (Rice \etal 2003; \S 5). 

For these reasons, we have therefore identified systems as IRAC Class II or III (Table 1),
based on the presence of significant excesses over photospheric levels 
{\em in the IRAC bands}.  This means that the SED
classifications differ in a few cases from those adopted by KH95.  Specifically, unless there
is a significant disk excess at 8 $\mu$m, we assign the object to IRAC Class III.
In the case of CoKu Tau 4, even though it was assigned Class II by KH95, and
has a small excess in the [8] band, we assign it to IRAC Class III as well, 
because it is much closer in its colors to the other Class III systems.

We find that the break in the distribution of colors
at around [5.8] - [8] $\sim 0.3-0.4$ corresponds accurately to the distinction between
Classical T Tauri Stars (CTTS) and Weak Emission T Tauri Stars (WTTS) (Herbig \& Bell 1988;
White \& Basri 2003).  As discussed in more detail in \S 6, the CTTS exhibit disk accretion
onto the central star, while the WTTS do not, and detectable accretion is 
generally accompanied by warm dust emission from the inner disk.  This distinction is 
clear in our IRAC data.  All of the CTTS (solid circles) are Class II, but not all of the 
WTTS (open circles) have previously been identified as Class III, i.e.  some WTTS 
have outer disks and longer-wavelength excess dust emission.  
The prime example of this is CoKu Tau 4, which is a WTTS with a small 
H$\alpha$ equivalent width of around 2 \AA\ (Cohen \& Kuhi 1979), 
indicating that there is very little if any accretion onto the central star.
In this case, the emission class is more consistent with the assignment of IRAC SED
Class (III) than with the previous Class II assignment by KH95.
In contrast, TW Hya is a CTTS, accreting from its disk, with strong 
H$\alpha$ emission and significant ultraviolet continuum excess radiation 
(Calvet \etal 2002; Muzerolle \etal 2000; Alencar \& Batalha 2002); 
thus, it is not surprising that TW Hya has a much redder [5.8] - [8] color than CoKu Tau 4. 

In general, the CTTS/II fall in a well-defined area of the color-color diagram.
The only exception, outside of TW Hya as described above, is 04301+2608.
This object has $[3.6] - [4.5] = 0.33$, typical of other disk systems, but a very
red long wavelength color,
$[5.8] - [8] = 1.67$, consistent with the relatively bright IRAS fluxes observed
(see \S 5).  Brice\~no \etal (2002) noted the very low apparent luminosity
of 04301+2608, given its M0 spectral type, and suggested that it might be an edge-on disk system
with substantial contributions of scattered light at shorter wavelengths.

The open squares and triangles are objects identified as Class I systems by KH95, plus
DG Tau B, which was not assigned a type by KH95.  DG Tau B is extremely red in
IRAC colors ($[3.5] - [4.5] = 1.61$; $[5.8] - [8] = 1.04$), 
and exhibits an extended bipolar nebulosity with a dark lane running
across the middle seen at optical and near-infrared wavelengths similar to that observed for
other Class I systems (e.g., Stapelfeldt \etal 1997; Padgett \etal 1999).
Because at least some Class 0/I objects show extended nebulosities, 
we show photometry for Class I objects for both the standard aperture radius of 5 pixels
(6 arcsec, corresponding to 840 AU at the 140 pc distance of Taurus), plus the standard
aperture corrections noted above for a point source, as well as a ``large aperture''
measurement with an aperture of 10 pixels (12 arcsec, 1680 AU).  The results 
for the differing apertures are almost identical except for 
IRAS 04368+2557, which is the driving source for the molecular outflow
in L1527.  This system is classified by Andr\'e et al. (2000) as a borderline Class 0/I
system, whereas Chen \etal (1995) identify it as a Class 0 system.
The relatively blue color in [5.8] - [8] for this object
is likely due to the wings of the $10 \mu$m silicate feature
affecting the flux in the [8] band, with a long-wavelength cutoff of about $9.3 \mu$m.
IRAS 04368+2557 appears as a very large, extended nebulosity without a central point source
in all the IRAC bands.  These images will be discussed
in a forthcoming publication (Allen \etal 2005).

One object, HV Tau C, has colors almost equal to those of the Class I source
$04260+2642$, while exhibiting a large H$\alpha$ equivalent width consistent with
it being a CTTS (White \& Ghez 2001).  (HV Tau AB itself is a WTTS/Class III close binary.) 
Optical and near-infrared imaging of HV Tau C shows a dark lane consistent with
absorption in a flattened circumstellar disk seen nearly edge-on (Monin \& Bouvier 2000; Stapelfeldt
\etal 2003).  However, Stapelfeldt et al. (2003) concluded that a dusty envelope appears to
be required in addition to a disk in order to explain the observations.  This suggests
that HV Tau C might actually be a Class I source, with a relatively tenuous infalling
envelope in addition to a circumstellar disk (see \S 6).

The crosses in Figure 1 denote systems with spectral types later than M6, and are thus
identified as probable brown dwarfs (Luhman 1999).  
Specifically, these objects are GM Tau (M6.5, White \& Basri 2003);
KPNO 4, 5, 6, 7 (M9.5,M7.5,M8.5,M8.25, respectively, Brice\~no \etal 2002);
and IRAS 04414+2506 (M7.25, Luhman 2004).
Of the six such objects observed so far,
four have clear IRAC excesses indicating the presence of disks -- and consistent
with the colors of disks of higher-mass T Tauri stars -- while one (KPNO 5)
clearly exhibits photospheric colors.  
The small [5.8] - [8] excess exhibited by the M9.5 star KPNO4 should 
be discounted because of large photometric errors for this faint object (Table 1).  Although
Brice\~no \etal (2002) estimated a large H$\alpha$ equivalent width suggestive of disk accretion
in KPNO4, a more recent high-resolution study (Muzerolle \etal 2003) indicates 
a much lower equivalent width, with marginal evidence of accretion from the H$\alpha$ profile.  
This system probably is not accreting from a circumstellar disk.  

In Figure 1 we also show the reddening vectors for a Vega-like spectrum
and $A_V = 30$, estimated from interpolating the Mathis (1990) extinction law,
applying this law to a model of the photospheric emission of Vega, and
integrating the resultant spectrum over the IRAC bandpasses.
The effect of circumstellar disk emission is to distribute the sources
redward in both colors; in contrast, circumstellar extinction is predicted to move sources
much more vertically, or even slightly blueward in the [5.8] - [8] color, due to
the effect of the 10 $\mu$m silicate feature on the IRAC [8] band as mentioned
above.  In contrast, systems with strong disk emission tend to be much redder in
[5.8] - [8]; many CTTS exhibit silicate features in emission,
producing a color shift distinct from circumstellar absorption.  Thus it appears that
the [3.6] - [4.5] vs. [5.8] - [8] diagram can be especially useful 
for distinguishing disk emission from the effects of interstellar absorption.  

\subsection{Other color-color diagrams}

Figure 2 shows the [3.6] - [4.5] vs. [4.5] - [5.8] diagram.  The general distribution of
points into the three main groups is more or less the same as in Figure 1, although the
division between C/Class II and W/Class III systems is less well defined.  Such a diagram
may be useful in many star-forming regions, where the sensitivity in band [8] 
is reduced by bright extended emission attributed to Polycyclic aromatic
Hydrocarbons (PAH).

Figure 3 shows other color-color diagrams using the shortest IRAC bands, which have
the highest sensitivity and are least affected by PAH emission, with near-infrared colors
taken from the 2MASS point source catalog.  If the true reddening vectors are reasonably
well-represented by the values adopted here, then it appears that combinations of deep 
near-infrared measurements and IRAC data can be used to distinguish between stars with
and without circumstellar dust emission (i.e., between Class III and Class II/I/0 systems).
The $H - $~[3.6] vs. [3.6]~-~[4.5] diagram appears to provide a particularly clean break,
at least for systems similar to those of the Taurus star-forming region.   

D'Alessio \etal (1999) used the ground-based and IRAS data listed in KH95 to construct
a ``median'' spectral energy distribution or SED for Taurus Class II stars of spectral
type K5-M2.  Because many Taurus stars in this spectral type range are yet to be observed
with IRAC, we defer the construction of a similar median SED to a later paper.  Here we
make a preliminary estimate of the consistency of the median SED with the observed colors
of the sample so far.  We do this by taking the fluxes of the median SED 
from D'Alessio \etal (1999) and interpolating linearly in log flux vs. log wavelength,
and then convolving the resulting spectrum with the predicted transmissions of the IRAC
bands (Fazio \etal 2004)  Because the ground-based data are sparsely sampled in the IRAC range
(most objects have only K,L, and N magnitudes, with few M magnitudes), this procedure is
relatively crude, but serves as an initial consistency check.  We thus estimate the following
IRAC colors for the median SED; [3.6] - [4.5] $= 0.40$, [4.5] - [5.8] $= 0.52$,
and [5.8] - [8] $= 0.83$.  As can be seen from the figures, these colors lie near the middle
of the distribution of CTTS or Class II colors, indicating reasonable agreement, though
the predicted [4.5] - [5.8] color is a bit redder than typical. 

\section{Comparison with models}

Figure 4 compares two of the main IRAC color-color diagrams with models for Class I and II systems.
The disk models shown are from the most recent calculations by D'Alessio et al. (2005b),
which include dust settling in the upper disk layers, and a more refined
treatment of the emission of the wall at the dust sublimation radius,
which takes into account the optically thin layers of the wall atmosphere.
The models are coded by color for the logarithm of the disk mass accretion rate
in units of solar masses per year, with $\mdot$ = -7 (blue), -8 (green), and -9 (red). 
At a fixed accretion rate, models are shown for two values of the 
inclination angle, cos $i$ = 0.5 and 0.86, 
and values of the dust depletion parameter $\epsilon$ = 1, 0.1, 0.01, and
0.001 (see D'Alessio et al. 2005b).\footnote{The parameter $\epsilon$ denotes
the factor by which the dust in upper layers of the disk is depleted from
the standard value.}  

The disk models reproduce the range of observed IRAC colors very well at accretion
rates spanning the expected values for Taurus systems (e.g., Gullbring \etal 1998,
Hartmann \etal 1998).  The current IRAC sample has only a modest number of objects
in common with the Gullbring \etal and Hartmann \etal accretion rate measurements;
therefore, we cannot say with confidence whether the detailed colors of individual
objects are consistent with model predictions given their measured mass accretion rates.
We will make a more detailed test of the model predictions when the
Taurus survey data are complete.

The Class I models are taken from Allen et al. (2004) for two luminosities,
 1 $\lsun$ (magenta lines) and 0.1 $\lsun$ (red lines), and two values 
of the centrifugal radius $R_c$ of 50 AU (solid line) and 300 AU (dashed line);
Values for the density scaling
parameter $\rho_1$ increase from bottom to top along a given line, with
markers (solid squares) at values of log $\rho_1$ = -14, -13.5, and -13.
\footnote{$R_c$ is the outer radius at which envelope material falls onto the
disk due to the angular momentum barrier; $\rho_1$ is the density the infalling envelope
would have at 1 AU if $R_c \rightarrow 0$ (see Kenyon, Calvet, \& Hartmann 1993 
for further discussion).  For a mass infall rate of $10^{-5} \msunyr$ to a $1 \msun$ protostar,
$\rho_1 \sim 5 \times 10^{-14} {\rm g\, cm^{-3}}$; $\rho_1 \propto \mdot M_*^{-1/2}$.}
These parameters have been found to span the characteristic
range needed to explain Taurus Class I SEDs including IRAS fluxes 
(Kenyon, Calvet, \& Hartmann 1993 $=$ KCH93). 

The models predict that objects with low mass infall rates and/or large centrifugal radii 
will have colors similar to that of ``pure disk'' systems. 
While the protostellar envelope models generally span the range of observed colors
for reasonable parameters, it is difficult to match the detailed parameters estimated
by KCH93 for individual objects with these models.  The reason is that
inferred parameters are very sensitive to the assumed geometry, as discussed further
in \S 6.

\section{Spectral Energy Distributions}

Figures 5 et seq. display SEDs for our program sample.  
The SEDS are constructed from the IRAC data, 2MASS JHK photometry, 
and when available, other photometry listed in KH95.  In many panels we show the 
median CTTS SED of D'Alessio \etal (1999), approximately scaled by eye
to the observed infrared fluxes, along with the long-wavelength colors of the K7 WTTS/
Class III star V819 Tau.
Our purpose in including these comparison SEDs is merely to provide a standard reference
in most panels for making rough intercomparisons between systems, without making extinction
corrections.  There has been no attempt to make detailed fits, which would require careful 
analysis of spectral types and optical colors to make the necessary reddening 
corrections.

In general, there is good agreement with ground-based data.
The Class I systems show a variety of behavior.  The most heavily-reddened systems show
evidence of silicate absorption, as expected.  
Among the Class II systems, many exhibit disk SEDs comparable to that of the median
SED, such as CX Tau, FV Tau/c, DF Tau, IQ Tau, DS Tau,
etc.  A few objects exhibit disk emission that falls off faster to longer wavelengths
than the median SED (e.g., CIDA 8, 11, 12, 14, DK Tau 1, and possibly JH 223).
DR Tau shows a very large excess consistent with strong accretion heating (e.g., 
Bertout, Basri, \& Bouvier 1988).
One peculiar result is that the apparent 2MASS magnitudes for CIDA 9 are extremely inconsistent
with the JHK magnitudes from KH95; the latter are much more consistent with the IRAC colors.
UZ Tau W (UZ Tau Ba $+$ Bb; White \& Ghez 2001) also poses a problem for analysis, because
the 2MASS H and K$_s$ magnitudes are about 0.7 and 0.8 mag brighter, respectively, than those 
listed by KH95 (see Figure 6), which produces a very large uncertainty in assessing the level 
of infrared excess.  The KH95 magnitudes are more consistent with the
White \& Ghez (2001) K magnitude for UZ Tau W as well as with the IRAC results.
UZ Tau EW, with a separation of 3.5 arcsec, is poorly resolved by
2MASS, which may be the source of the discrepancy.

GM Aur is a CTTS with a small IRAC excess but much larger excesses at 
$\lambda \geq 10 \mu$m (Rice \etal 2003), similar in many respects to the 
CTTS TW Hya and the WTTS CoKu Tau 4, as discussed above.  
KH95 identified DI Tau as a Class II object, citing a $10 \mu$m excess with a very large error,
whereas we find no evidence for excesses out to band [8], and thus identify it
as Class III.
Meyer \etal (1997) detected no $10 \mu$m excess based on detailed ground-based data;
IRAS emission in the region of this source is dominated by the nearby CTTS DH Tau,
and thus whether any excess is present at long wavelengths is currently uncertain. 

The WTTS systems LkHa332 G1 and G2 have no IRAC excesses; thus we identify both as Class III
systems, in contrast to the Class II given by KH95 for G1. The
IRAS fluxes in the region are consistent with arising entirely from LkHa 332 $=$ V955 Tau.
We find excess emission from UZ Tau w, in contrast to the Class III identification in KH95,
but consistent with the large H$\alpha$ equivalent widths of this close binary system found
by Hartigan \& Kenyon (2003).

Our small sample of brown dwarfs with disks have colors similar to those of higher-mass CTTS.
This is undoubtedly because the disk emission is optically thick, and geometrical
properties of the disks are relatively unimportant in the IRAC range, i.e. the disks are thought
to be geometrically flat except possibly at the inner edge of the dust disk (see, e.g. Dullemond
Dominik \& Natta 2001; Luhman \etal 2005).

\section{Discussion}

One of the purposes of conducting this survey of Taurus is to investigate how reliably
objects can be classified using IRAC colors by reporting results for well-known,
independently-classified systems.  As mentioned in \S 3.1, 
the major system of infrared classification
depends upon an infrared spectral index, typically measured from about $1 - 2 \mu$m to about 
$10 - 20 \mu$m (Lada \& Wilking 1984; Adams \etal 1987).  
Plotted as $\log \lambda F_{\lambda}$ vs. $\log \lambda$, Class 0/I systems have positive spectral
indices, indicating substantial circumstellar envelopes and are usually thought to be protostars
(Adams \etal 1987, Andr\'e, Ward-Thompson, \& Barsony 1993).  Class II sources have negative slopes, 
but far less steep than that of a Rayleigh-Jeans spectrum, and are identified as systems with
dusty circumstellar disks (Lynden-Bell \& Pringle 1974; Adams \etal 1987; Kenyon \& Hartmann 1987);
and Class III systems have steep negative slopes consistent with those expected for a (reddened) 
stellar photosphere, essentially a Rayleigh-Jeans slope (Lada \& Wilking 1984; Adams \etal
1987).  

The other system for classifying low-mass pre-main sequence stars, as previously
mentioned in \S 3.1, depends upon spectroscopic criteria, introduced
by Herbig \& Bell (1988) and refined by White \& Basri (2003).  Using criteria
based on H$\alpha$ emission - usually a spectral-type-dependence of equivalent width,
correlated in some cases by velocity width - one may divide systems into 
``Classical'' T Tauri stars or CTTS vs. ``Weak'' T Tauri stars or WTTS.  The WTTS have
relatively weak H$\alpha$ equivalent widths, and narrow H$\alpha$ line profiles
produced by chromospheric activity; the CTTS have larger H$\alpha$ emission
and broader profiles thought to be produced by disk accretion (see discussion in Hartmann 1998).

We have shown that IRAC colors can be used to classify systems accurately between
CTTS and WTTS.  This is not surprising, given that accretion onto the central
star has been known to correlate with infrared excess emission in the $L$
photometric band (Hartigan \etal 1990).  
Basically, it appears that all systems with detectable gaseous accretion onto the
central star (through {\em either} ultraviolet continuum excesses or broad H$\alpha$ profiles)
have enough small dust particles carried along with the accreting gas to provide significant
near-infrared excess emission.  Whether a system should be identified as Class II or
Class III is more complicated, depending upon whether the disk has an inner hole.
We have skirted this issue of nomenclature in this paper by defining systems based on
the presence or absence of excesses in the IRAC bands.
 
The agreement between the observed Taurus colors and the current disk models is encouraging.
These model predictions are very similar to those presented in Allen \etal (2004) and Megeath
\etal (2004).  The Taurus results suggest that the methods used in the above papers are
adequate for identifying Class II/CTTS in star-forming regions, with appropriate allowances
for extinction.  

The disk models predict that the IRAC colors get redder as the mass accretion rate increases,
although inclination and dust properties also contribute in a complicated manner.
A preliminary investigation of [3.6] - [4.5] vs. mass accretion rates from the literature
is not conclusive, given the small number of systems with accretion rates in our current
sample; we will revisit this question when the Taurus survey is complete.

The disk models of Whitney \etal (2004) do not strongly populate the region of CTTS/Class II
sources found observationally here.  As Whitney \etal note, this is likely because
they have few models appropriate for low-mass stars.  The protostellar envelope models
of Whitney \etal similarly do not show colors corresponding to L1527, but this may be
due to the relatively small aperture we used for the photometry of this object
considering the spatial extension of its emission (Allen \etal 2005).

We also find that our small sample of Taurus Class I systems tend to divide into two
comparably-sized groups; one that is very much redder than Class II systems, and another
that exhibits IRAC colors which are similar to the reddest Class II objects (\S 3.1).
Using the Class 0-I criteria to distinguish protostellar systems from Class II/T Tauri systems
implicitly depends on protostellar envelopes having significant opacity in the IRAC bands,
especially along the line of sight to the central (proto)star; otherwise the
envelope can neither absorb much light or produce much emission in the IRAC bands.
The extinction produced by an infalling protostellar envelope depends upon the infall rate, 
the angular momentum of the infalling
material, the size of envelope holes driven by outflows, and departures from sphericity
of the initial protostellar core, along with the system inclination to
the line of sight (Adams \etal 1987; KCH93; Calvet \etal 1994; Hartmann, Boss, \& Calvet 1996;
Whitney \etal 2003).  

It is instructive to make a simple estimate of envelope optical depth in the IRAC region,
ignoring complicating geometric factors for the moment.
Assuming a spherical envelope in steady free-fall at mass infall rate $\dot{M}$
towards a central protostar of mass $M_*$, the radial mass column density is
\begin{equation}
\int^\infty_{r_0} dr \, \rho = {\dot{M} \over 2 \pi (2 G M_* )^{1/2}{r_0}^{1/2}}\,,
\end{equation}
where $r_0$ is the inner radius.  If we identify $r_0$ as comparable to the centrifugal
radius, i.e. roughly the radius at which material falls out onto the disk, we can then
make an estimate of the envelope optical depth as a function of this parameter.
Using Draine \& Lee (1984) opacities at $\lambda = 3.4 \mu$m, we find numerically

\begin{equation}
\tau(3.4 \mu{\rm m}) \sim
1 \, \left ( {\dot{M} \over 4 \times 10^{-6} \msunyr } \right )\,
\left ( {M \over 0.5 \msun} \right )^{-1/2} \,
\left ( {r_0 \over 100 {\rm AU}} \right )^{-1/2}\,.
\end{equation}

The fiducial value $\dot{M} = 4 \times 10^{-6} \msunyr$ is the typical infall rate
for Taurus Class I sources deduced by KCH93, based mostly on the analysis
of longer-wavelength IRAS fluxes.  KCH93 also estimated centrifugal
radii (comparable to $r_0$) of 70-300 AU for most systems, although with considerable
uncertainty.  For lower values of envelope column density ( lower values of $\dot{M} r_o^{-1/2}$),
the envelope will be essentially transparent at IRAC wavelengths, while higher values of
envelope column density will result in potentially significant effects of the envelope
in extincting the central source and producing extra dust emission.  
Thus, it is not very surprising that (of our small sample), four of the Class 0/I sources
have IRAC colors significantly different from those of Class II (disk) systems
(04368$+$2557, 04365$+$2535, 04381$+$2540, DG Tau B) while three Class I systems exhibit
colors at the red end of the disk systems (04260$+$2642, 04248$+$2612, 04489$+$3042).
We suggest that many or most Class I systems with infall rates similar to
or larger than those typical of Taurus should exhibit distinct IRAC colors from those
of Class II systems.

The above estimate neglects the uncertainties that can arise in classifying individual
systems due to additional departures from sphericity in the envelope, such as bipolar
outflow holes or collapsing flattened envelopes (e.g., Hartmann \etal 1996).  
To expand upon this, we note that of the sources in common with
those modelled by KCH93, the reddest systems in [3.6] - [4.5]
(04368+2557, 04365+2535, and 04381+2540) had relatively high values of the density
scaling parameter $\rho_1$, consistent with the IRAC observations.
Conversely, the 04489+3042 system is bluer in [3.6] - [4.5] and had a low $\rho_1$
according to the modelling of KCH93.  The system 04248+2612 has relatively blue
colors, even though KCH93 found a high value of $\rho_1$ in their fitting.  However,
the fit of KCH93 adopted a large value of $R_c = 300$~AU and a very low viewing
inclination $i = 10^{\circ}$, and both parameter choices reduce the line-of-sight column
density in these rotating collapse models.  These results illustrate but
probably underestimate the sensitivity of the IRAC colors to protostellar envelope 
geometry, because they do not include bipolar outflow holes.  Viewing a highly embedded
protostellar source along an evacuated cavity could easily produce IRAC colors determined
entirely by the disk properties of the system.  

Alternatively, one might ask whether the Class I group with similar colors as
Class II systems might not simply be identified Class II as well.  Chiang \& Goldreich
(1999) suggested that many objects identified as Class I might really be Class II,
i.e. they disk systems seen more nearly edge-on; White \& Hillenbrand (2004) recently made
a similar suggestion.  A problem with this idea is that edge-on disks do not
necessarily, or even naturally, exhibit the same SEDs and colors as Class I envelope systems.
As discussed by D'Alessio \etal (1999) in some detail, because T Tauri disks are thought
to be very optically thick, the observed SED is extremely sensitive to the inclination
of observation.  As shown in Figure 13 of D'Alessio \etal (1999), which shows the predicted
spectrum of a disk model at different viewing angles, there is only a small
range of inclinations for which the SED is roughly comparable to that of Class I systems.
Slightly more pole-on, the SED looks like that of a normal Class II object; slightly more
edge-on, huge silicate absorption is predicted.  

In our sample of Class I systems, the group with nearly ``Class II colors'' does not 
exhibit the blue [5.8] - [8] color one would expect for deep silicate absorption, 
nor is such absorption apparent in ground-based data (e.g., Figure 5), and so these 
systems are unlikely to be explained as purely edge-on disks.  Note that the HV Tau C
system, whose image clearly shows absorption from a nearly edge-on disk, shows no
evidence for strong silicate absorption from the [5.8] - [8.0] color;
however, as noted in \S 3, Stapelfeldt \etal (2003) concluded that
an extended dusty envelope probably must be included along with a disk to explain
the near-infrared scattered light image, and this dust could in principle provide 
compensating silicate feature emission.  Stapelfeldt \etal showed that the infall
rate for such an envelope would be quite low, $\sim 10^{-7} \msunyr$, suggesting
that this object may be in transition between Class I and Class II. 

D'Alessio \etal (1999) also pointed out that edge-on Class II disk systems should appear
to be underluminous, by roughly an order of magnitude.  The three Class I systems
with colors comparable to the reddest Class II objects, 04260+2642, 04248+2612, and
04489+3042, have apparent bolometric luminosities of $0.09, 0.36$, and $0.3 \lsun$ respectively 
(KH95).  Hubble Space Telescope images of 04260+2642 indicate that it is an edge-on
disk system (Stapelfeldt 2005, personal communication), consistent with its relatively 
low luminosity (D'Alessio et al. 1999).  The other two objects have luminosities only
marginally below the median luminosity of Class I systems in Taurus (KH95).  Furthermore,
04248+2612 has a well-observed, extended near-infrared reflection nebulosity typical
of Class I systems (Lucas \& Roche 1997; Padgett et al. 1999;
Park \& Kenyon 2002), and NICMOS observations
at 1.1 $\mu$m using {\em Hubble Space Telescope} also show extended nebulosity around
04489+3042 in a pattern much more similar to that of envelope than edge-on disk systems
(L. Hartmann, unpublished).
We conclude that 04248+2612 and 04489+3042 are likely to be true Class I systems.

To make more progress in understanding the properties (or even existence) of protostellar
envelopes, inclusion of longer-wavelength data is crucial.  Comprehensive observations of
a range of Class I systems with IRAC, IRS, and MIPS should yield significant new insights
into protostellar evolution.

\section{Conclusions}

We have presented results from a survey of 82 Taurus pre-main sequence stars as observed
with the IRAC camera on {\em Spitzer} Space Telescope.  These data provide the
first indication of the presence or absence of infrared excesses 
in the 3-8 $\mu$m range for about half of our sample,
and add considerably to our knowledge of the SEDs of many more systems.  
The results generally confirm the 
color classification scheme used in previous IRAC studies of star-forming regions.  
The predictions
of disk and protostellar envelope models are consistent with the observed IRAC colors,
and suggest a relatively clear difference in between stars 
with and without optically-thick disks.
Among systems with disks, we find a very strong correlation of IRAC excesses with the
presence of accretion.  Our results should be of use for testing more detailed models
of disks and envelopes in pre-main sequence systems, and for interpreting IRAC
observations of more distant, less well-studied star-forming regions.

We thank the referee, Karl Stapelfeldt, for useful comments and for communicating
new results.
This work is based on observations made with the
{\em Spitzer} Space Telescope, which is operated by the
Jet Propulsion Laboratory, California Institute of Technology,
under NASA contract 1407.  This work was supported in
part through contract 1256790 issued by JPL/Caltech.
The research of L.H. and N.C. was supported in part by NASA grants NAG5-9670,
NAG5-13210, NAG5-10545, and grant AR-09524.01-A from the Space Telescope Science Institute.  
K.L. was supported by grant NAG5-11627 from the NASA Long-Term
Space Astrophysics program.

\facility{}

\clearpage

\begin{figure} 
\plotone{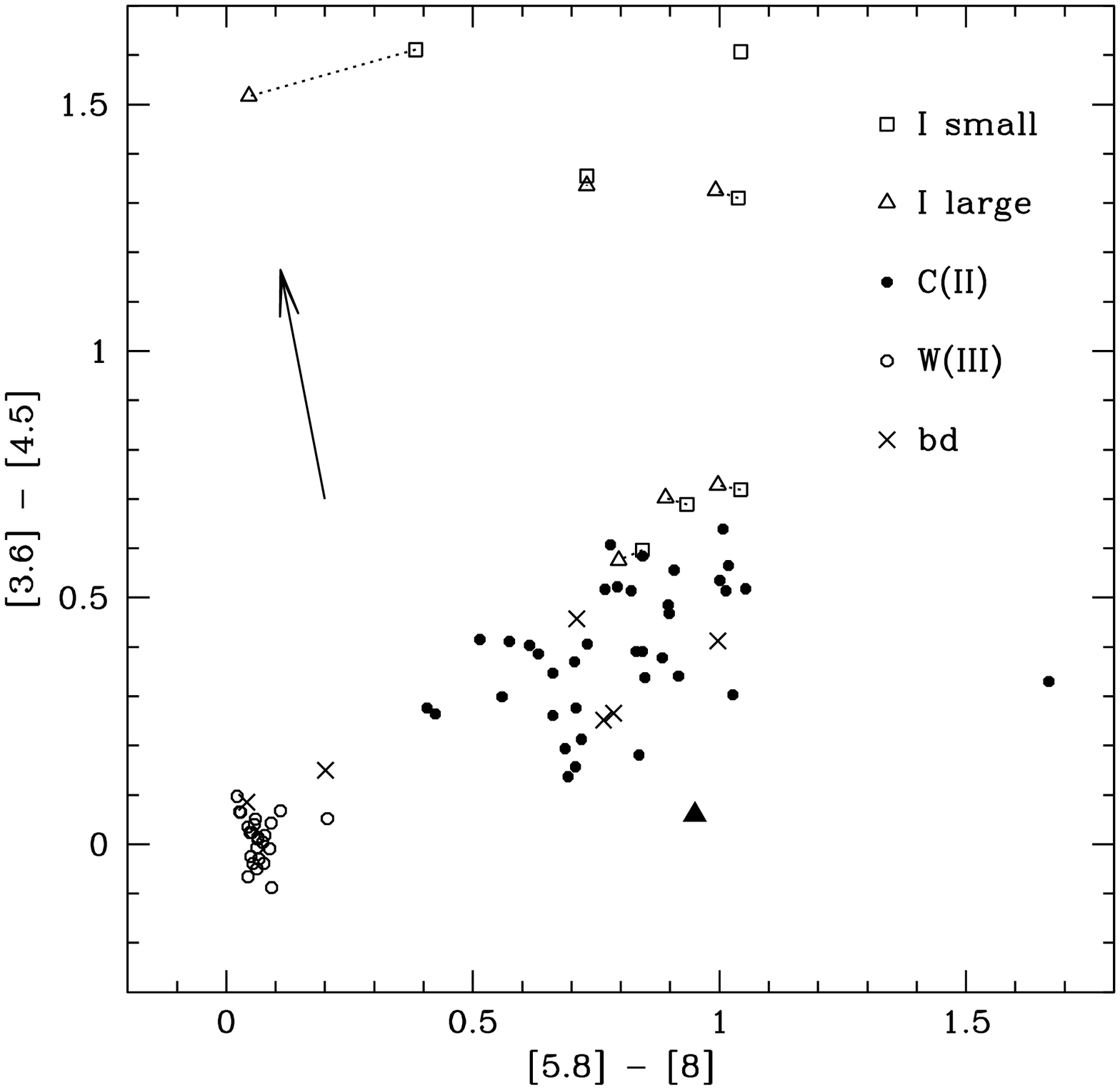}
\caption{Color-color plot for the four IRAC bands. Open circles are WTTS/Class III
objects, filled circles are CTTS/Class II systems, and open squares and triangles
are Class 0/I systems measured in large and small apertures (see text for explanation).
The crosses are brown dwarfs, and the solid triangle is TW Hya.  Reddening vectors
for $A_V = 30$ are shown for the case of a Vega-like spectrum 
and the reddening law listed by Mathis (1990).}
\end{figure}

\begin{figure} 
\plotone{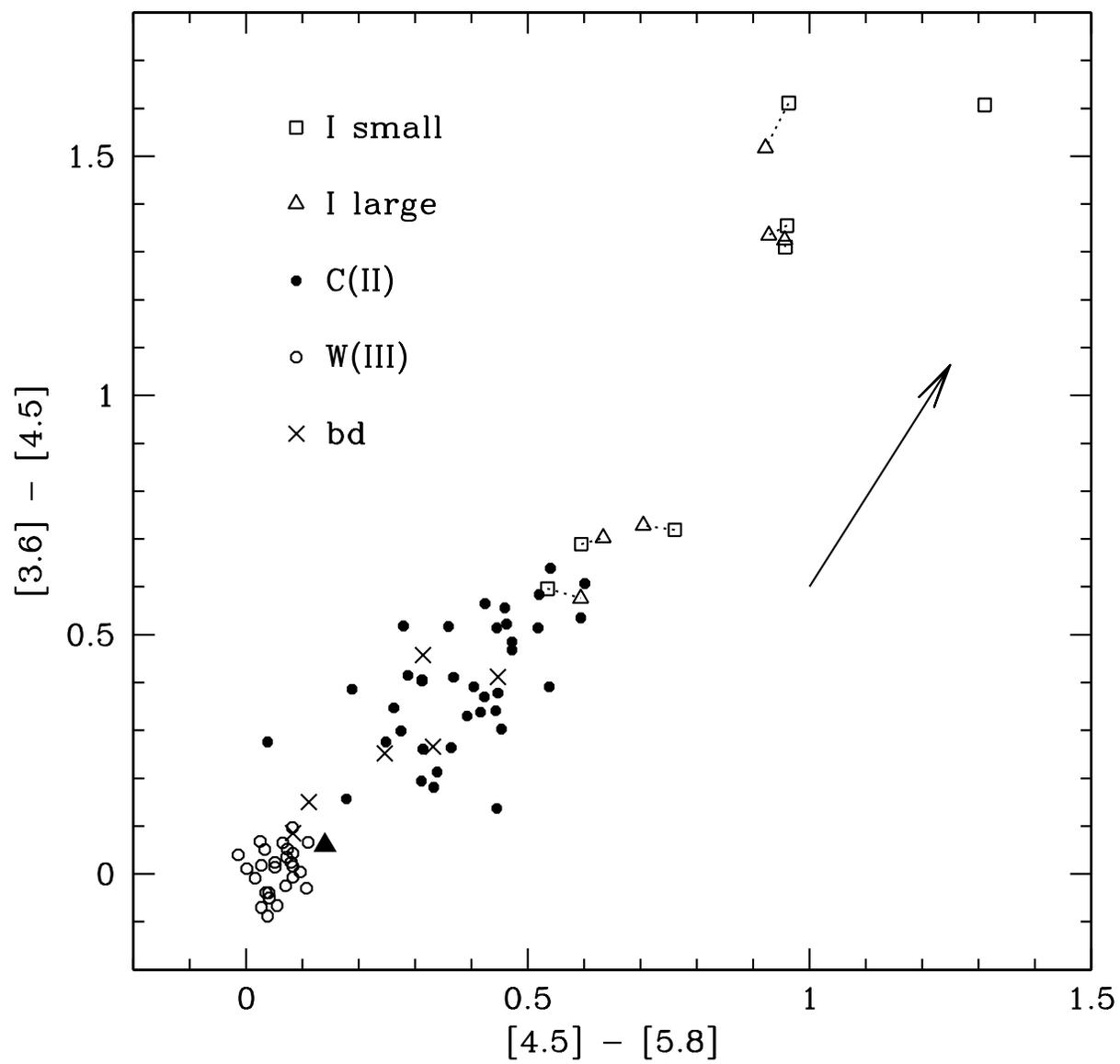}
\caption{Alternative IRAC color-color plot, eliminating the longest-wavelength band.
Symbols are the same as in Figure 1.}
\end{figure}

\begin{figure} 
\plottwo{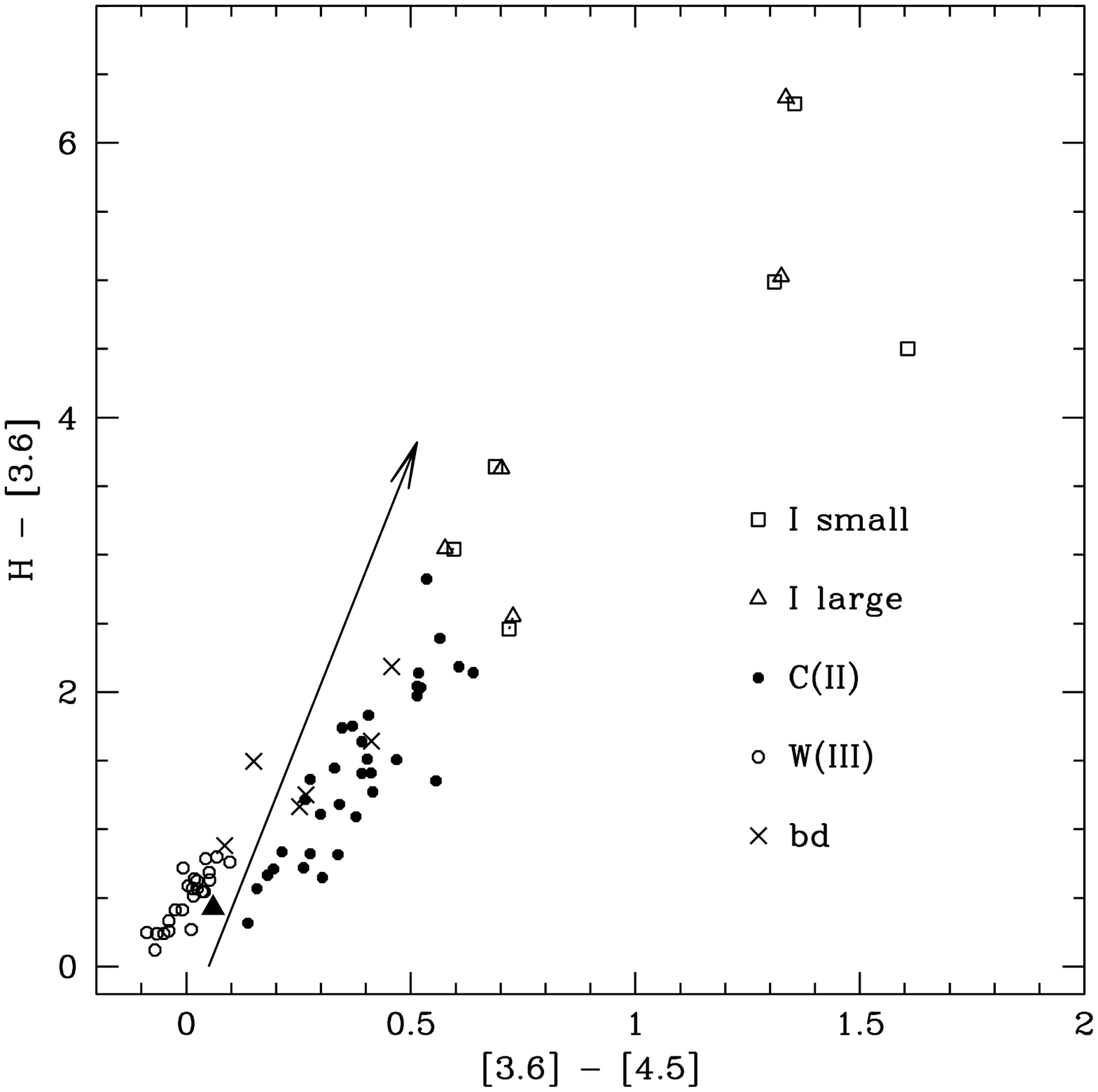}{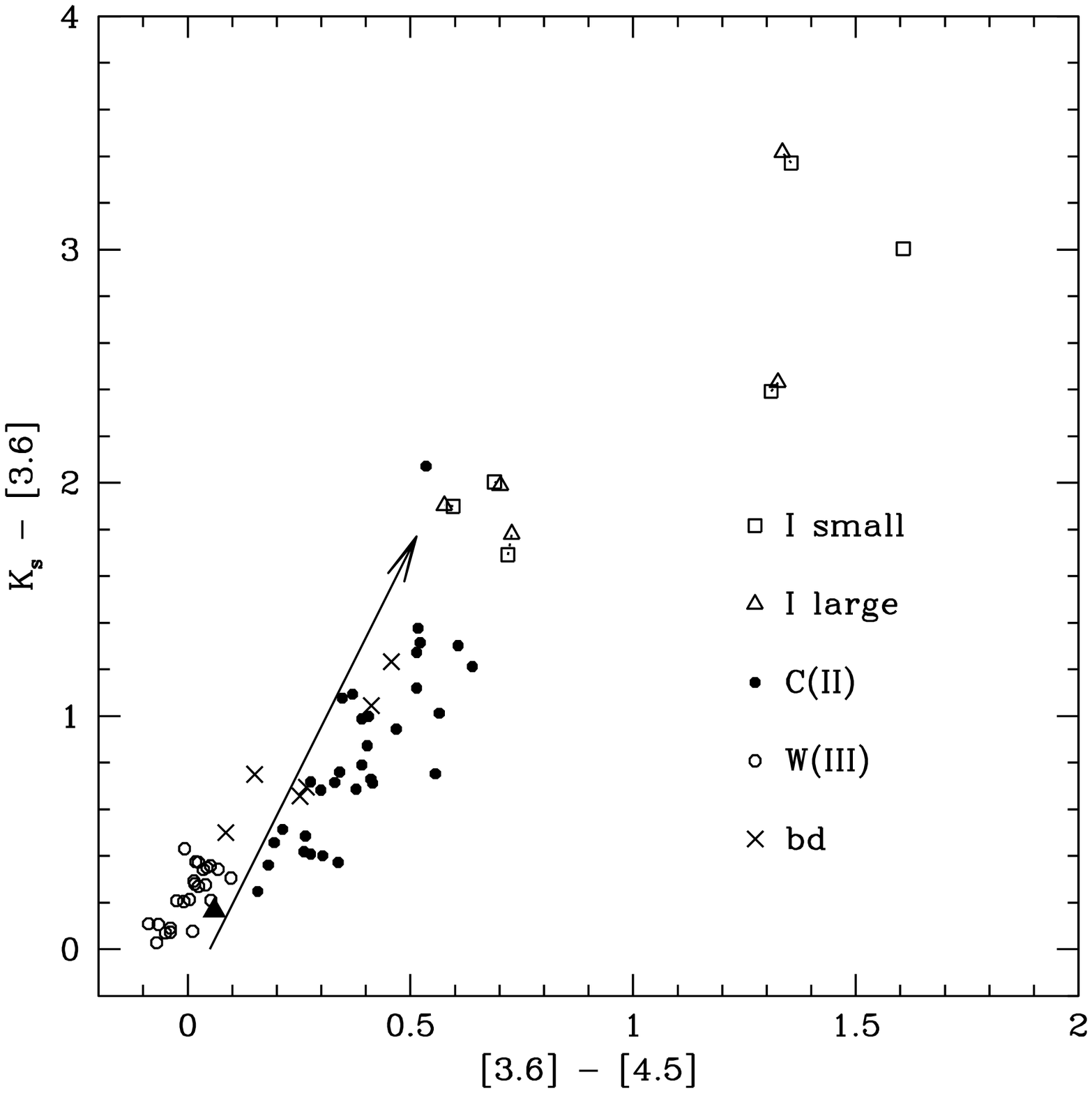}
\caption{Color-color plots using near-infrared 2MASS magnitudes and the
shortest-wavelength IRAC bands.}
\end{figure}

\begin{figure}
\plotone{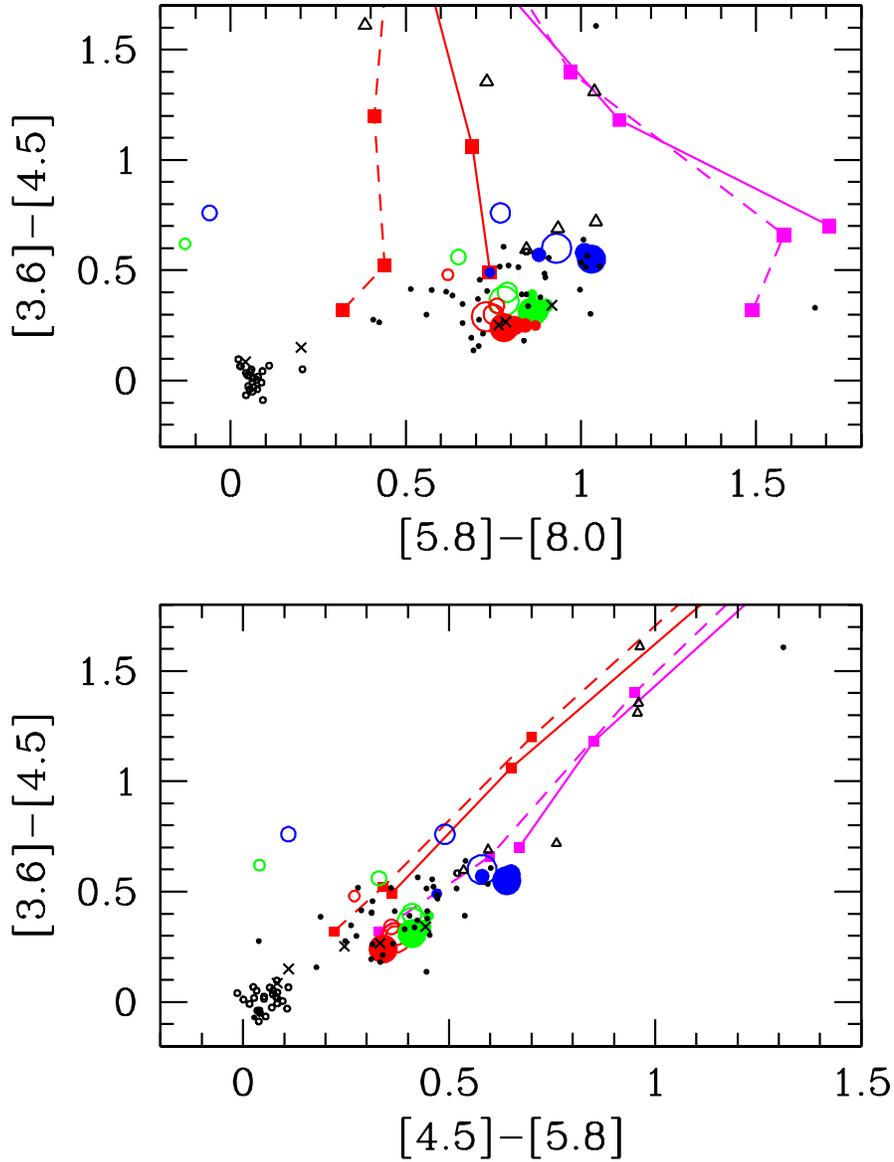}
\caption{Comparison of observed and model colors. The data are shown in
black, with symbols as in Figures 1 and 2. Class I models are taken from
Allen \etal (2004) for two luminosities, 1~$\lsun$ (magenta lines) and 
0.1~$\lsun$ (red lines), and two values of the centrifugal radius $R_c$ of 50
AU (solid line) and 300 AU (dashed line). Values for the density scaling
parameter $\rho_1$ increase from bottom to top along a given line, with
markers (solid squares) at values of log $\rho_1$ = -14, -13.5, and -13.
Disk model colors are taken from D'Alessio \etal (2005b) and are shown 
for values of the log of the mass accretion rate log
$\mdot$ = -7 (blue), -8 (green), and -9 (red). 
Disk models are shown for two values of the inclination angle, 
cos $i$ = 0.5 (solid circles) and 0.86 (open circles), 
and for four values of the depletion parameter, $\epsilon$ = 1, 0.1, 0.01, and
0.001 (D'Alessio \etal 2005b), 
labeled by symbol size; the smallest size corresponds to $\epsilon = 1$. }
\end{figure}

\begin{figure}
\plotone{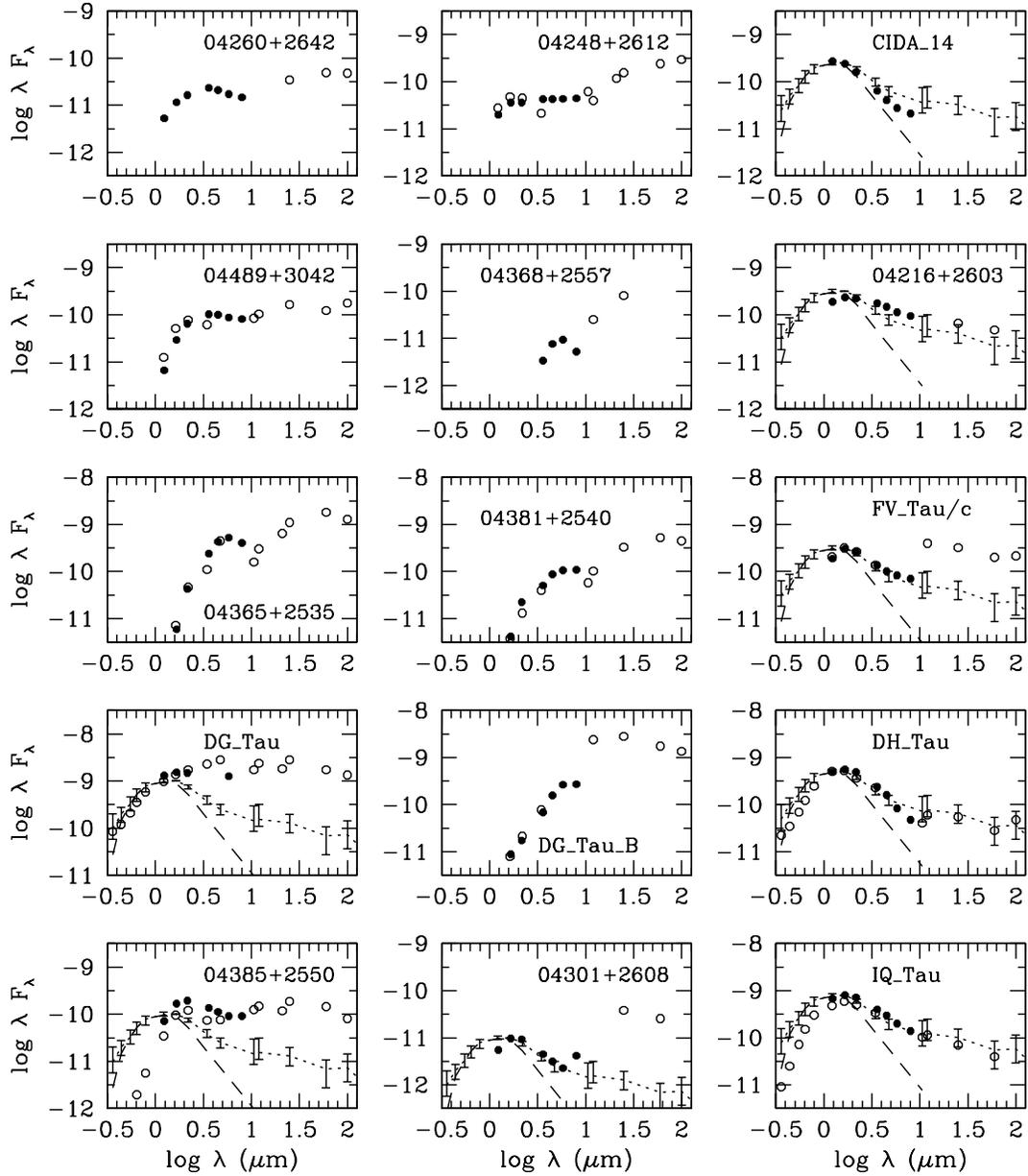}
\caption{SEDs of observed Taurus objects.  IRAC fluxes are combined with 
data from 2MASS (solid circles) and with
ground-based and IRAS data taken from Kenyon \& Hartmann (1995) (open circles).
For some objects we show the (dereddened) SED of the Class III star V819 Tau
(dashed line) and the median SED of Taurus Class II K7-M2 objects derived
by D'Alessio \etal (1999) (dotted line with error bars denoting the quartiles
of the distribution); this median SED illustrates the typical disk emission
in Taurus.}
\end{figure}

\begin{figure}
\plotone{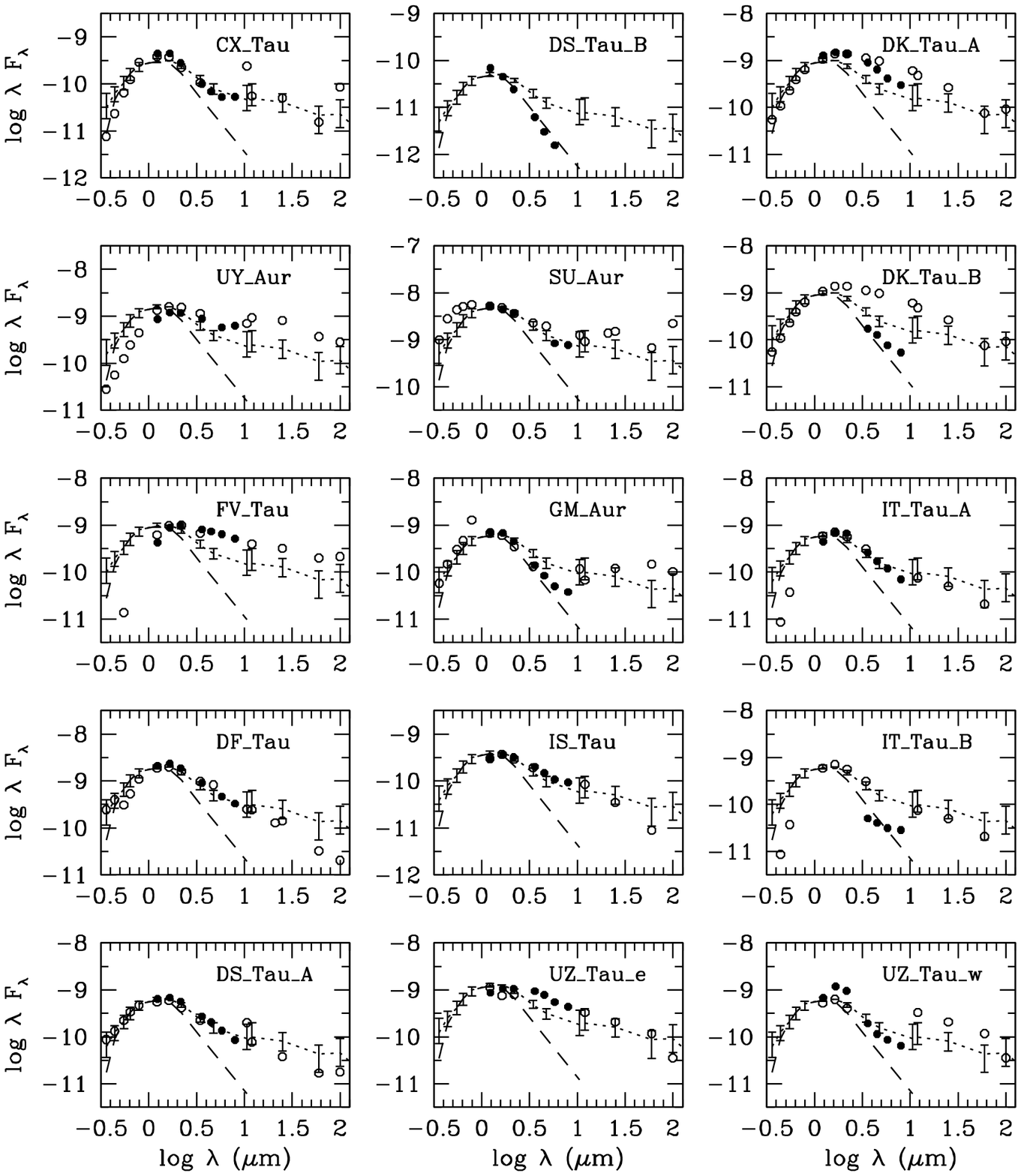}
\caption{Same as Figure 5.
}
\end{figure}

\begin{figure}
\plotone{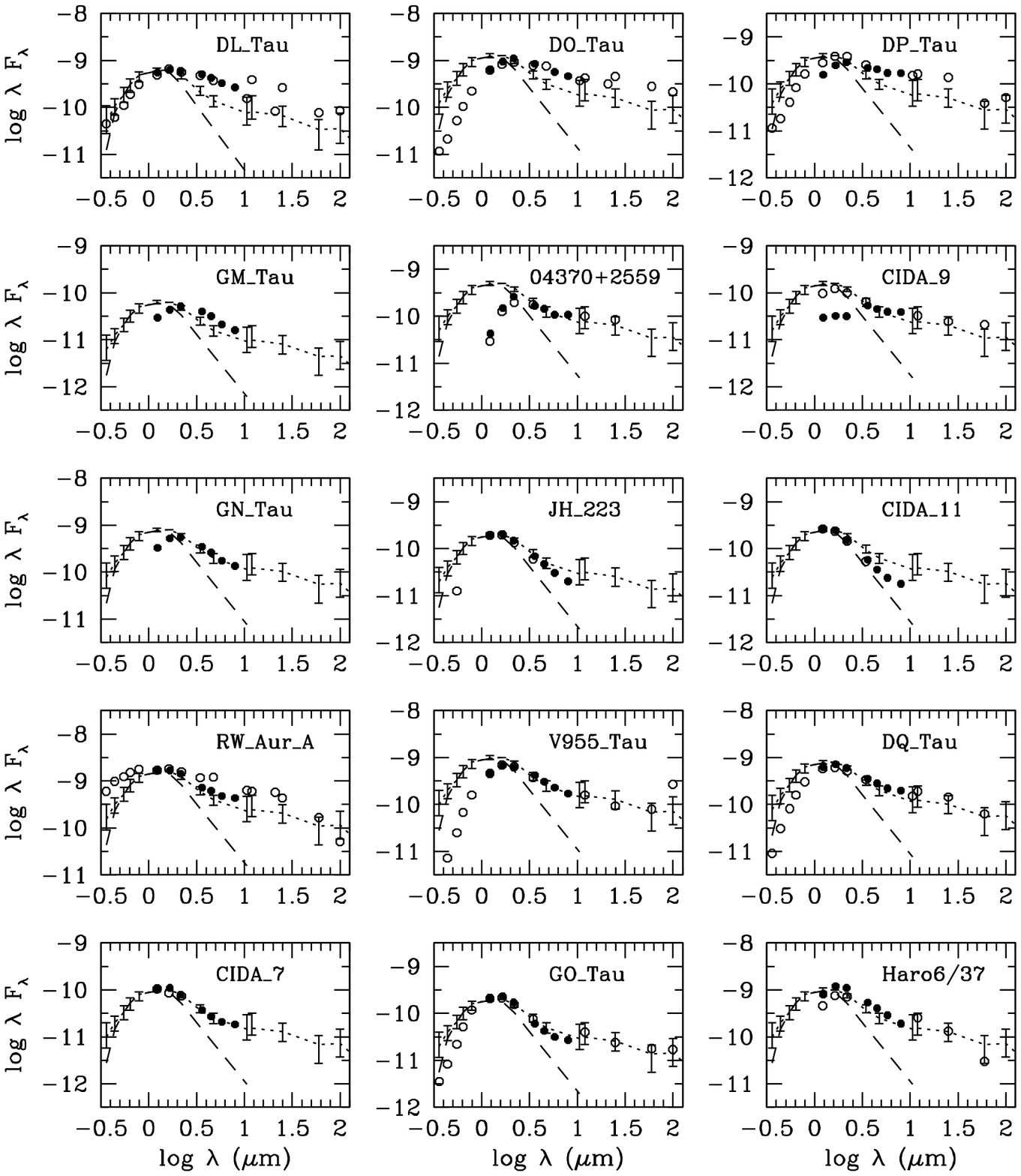}
\caption{Same as Figure 5.}
\end{figure}

\begin{figure}
\plotone{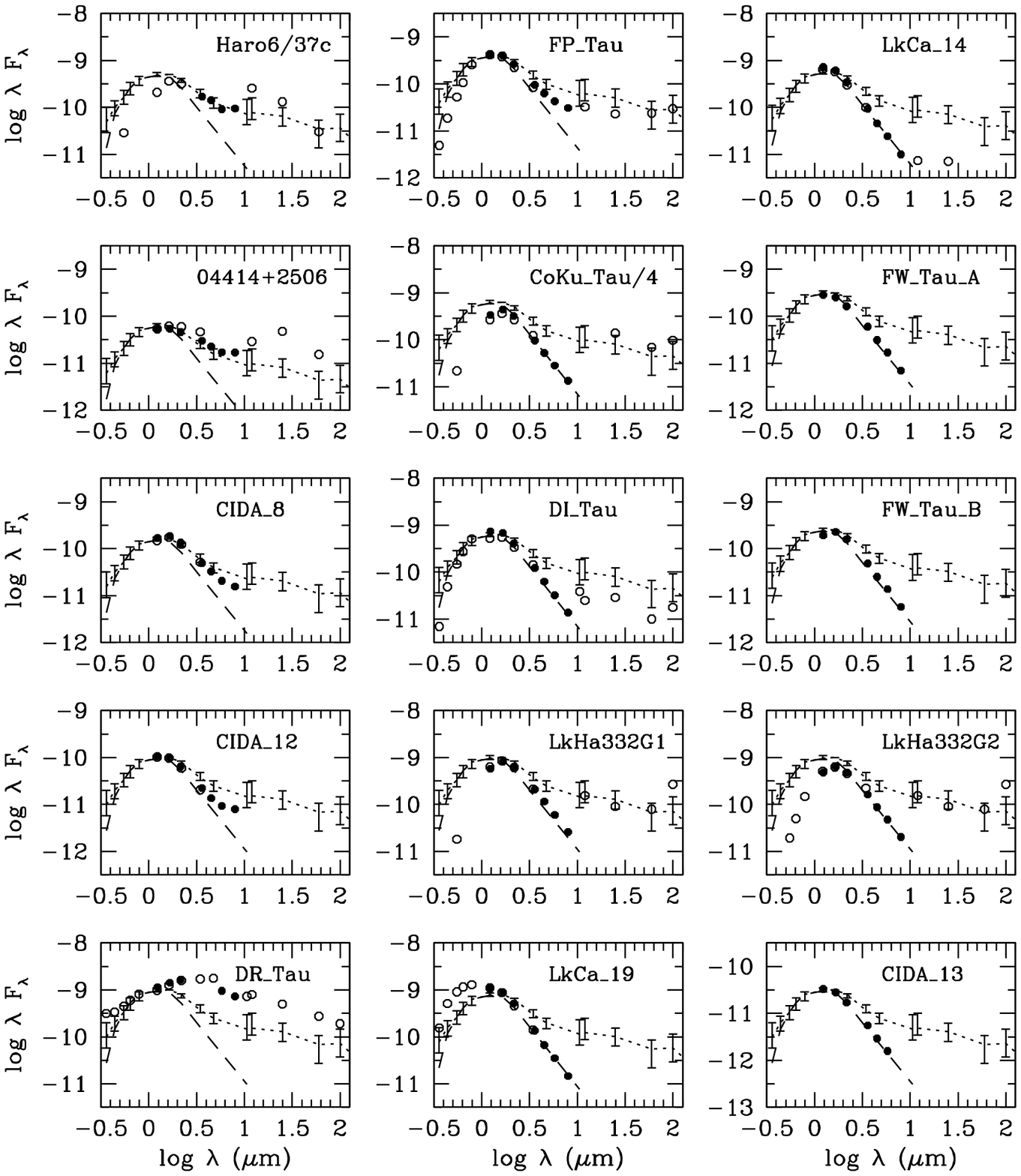}
\caption{Same as Figure 5.}
\end{figure}

\begin{figure}
\plotone{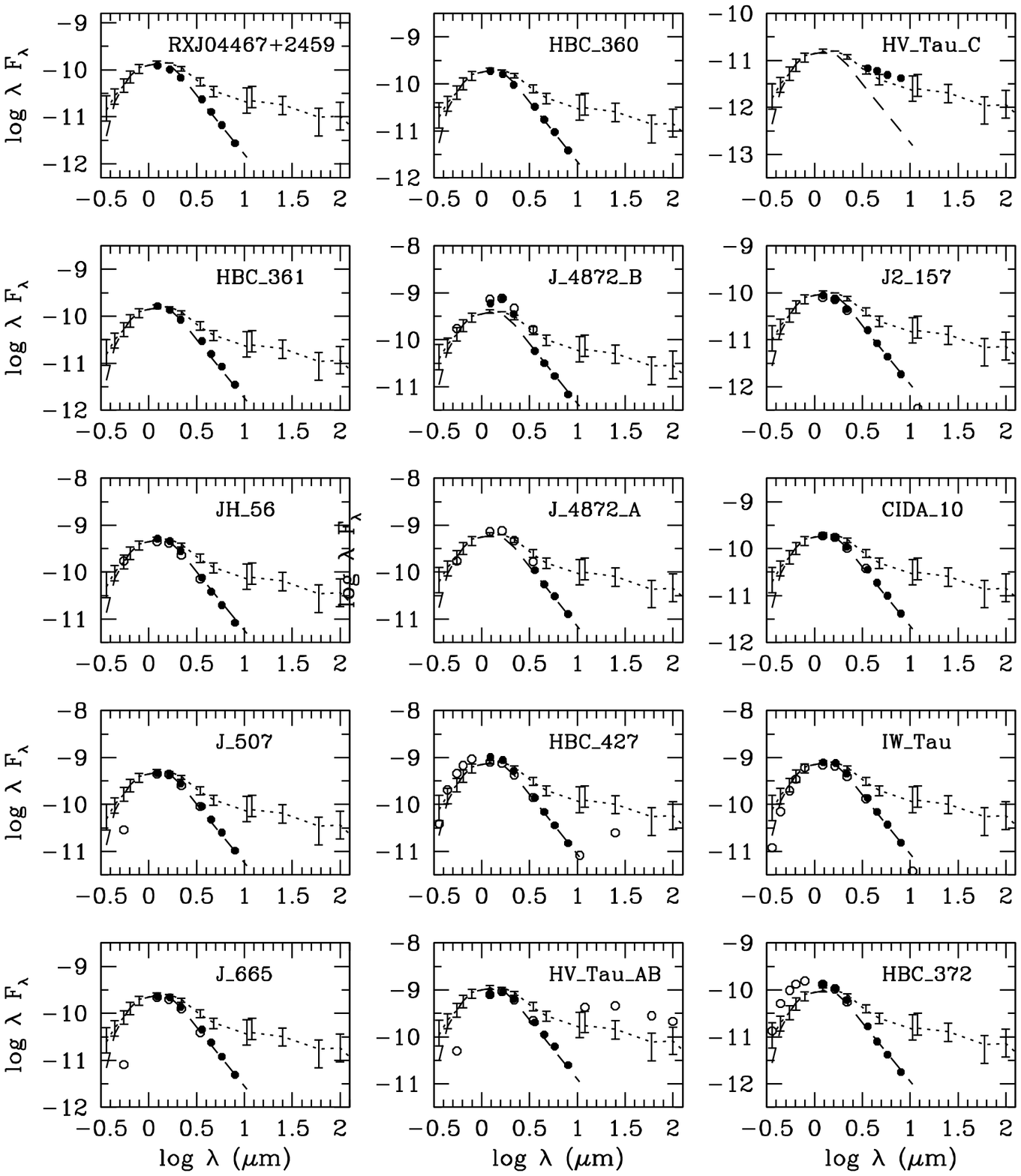}
\caption{Same as Figure 5.}
\end{figure}

\begin{figure}
\plotone{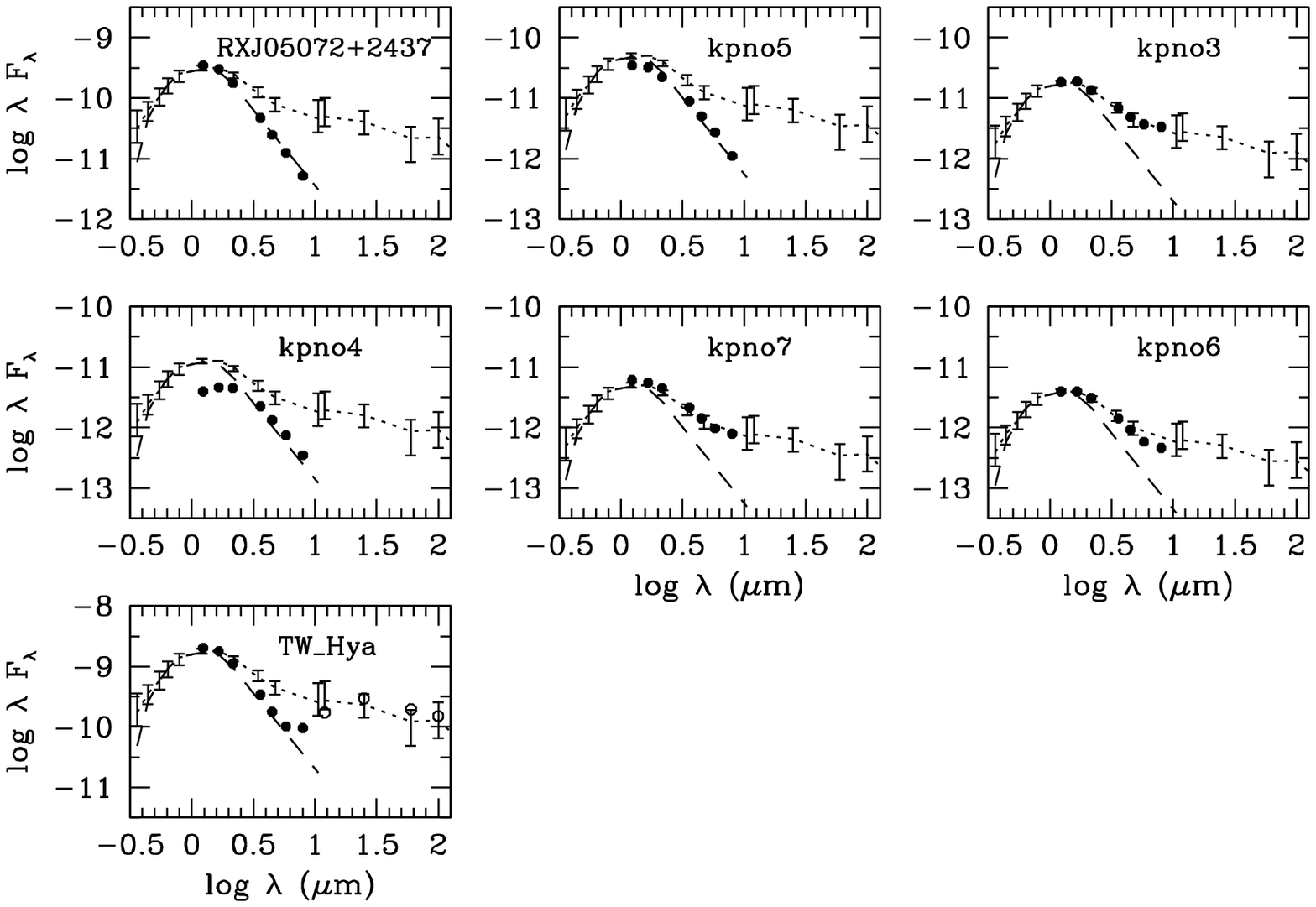}
\caption{Same as Figure 5.}
\end{figure}

\newpage

\pagestyle{empty}
\voffset 1.0in
\begin{deluxetable}{rrrrrrrrrrrrr} 
\rotate
\tabletypesize{\scriptsize}
\tablecolumns{13} 
\tablewidth{0pt}
\tablecaption{IRAC and 2MASS magnitudes} 
\tablehead{ 
\colhead{Name} & \colhead{2MASS ID}  & \colhead{RA}  & \colhead{Dec} & \colhead{J} & \colhead{H}   & \colhead{K$_s$}    
& \colhead{[3.6]}    & \colhead{[4.5]}   & \colhead{[5.8]}    & \colhead{[8]} & \colhead{JD}   & \colhead{IRAC} \\
\colhead{} & \colhead{}  & \colhead{(J2000)}  & \colhead{(J2000)} & \colhead{} & \colhead{}   & \colhead{}    
& \colhead{}    & \colhead{}   & \colhead{}    & \colhead{} & \colhead{$-$53000}   & \colhead{type}}
\startdata 
     04260+2642 & 04290498+2649073 & 67.27046 & 26.81819 & 14.680 & 13.025 & 11.883 &  9.984 $\pm$  0.011 &  9.388 $\pm$  0.040 &  8.853 $\pm$  0.008 &  8.009 $\pm$  0.008 & 71.676      & I \\
     04248+2612 & 04275730+2619183 & 66.98861 & 26.32129 & 13.235 & 11.795 & 11.026 &  9.335 $\pm$  0.019 &  8.616 $\pm$  0.014 &  7.855 $\pm$  0.007 &  6.812 $\pm$  0.006 & 71.676      & I \\
     04489+3042 & 04520668+3047175 & 73.02742 & 30.78775 & 14.426 & 12.021 & 10.383 &  8.379 $\pm$  0.011 &  7.690 $\pm$  0.012 &  7.095 $\pm$  0.008 &  6.161 $\pm$  0.012 & 49.355      & I \\
     04368+2557 & \nodata  & 69.97395 & 26.05224 & \nodata & \nodata & \nodata & 12.095 $\pm$  0.243 & 10.484 $\pm$  0.046 &  9.521 $\pm$  0.088 &  9.137 $\pm$  0.062 & 71.637     &  0 \\
     04365+2535 & 04393519+2541447 & 69.89638 & 25.69535 & 16.905 & 13.752 & 10.837 &  7.466 $\pm$  0.008 &  6.111 $\pm$  0.008 &  5.151 $\pm$  0.004 &  4.420 $\pm$  0.007 & 71.641      & I \\
     04381+2540 & 04411267+2546354 & 70.30240 & 25.77603 & 17.150 & 14.137 & 11.541 &  9.149 $\pm$  0.033 &  7.839 $\pm$  0.011 &  6.882 $\pm$  0.011 &  5.844 $\pm$  0.003 & 44.859      & I \\
         DG Tau & 04270469+2606163 & 66.76920 & 26.10399 &  8.691 &  7.722 &  6.992 & \nodata $\pm$ \nodata & \nodata $\pm$ \nodata &  4.191 $\pm$  0.004 & \nodata $\pm$ \nodata  & 71.684   & I/II \\
       DG Tau B & 04270266+2605304\tablenotemark{a} & 66.76040 & 26.09137 & 15.600 & 13.317 & 11.819 &  8.816 $\pm$  0.025 &  7.209 $\pm$  0.009 &  5.898 $\pm$  0.012 &  4.855 $\pm$  0.006 & 71.684& I/II \\
     04385+2550 & 04413882+2556267 & 70.41122 & 25.94040 & 11.849 & 10.123 &  9.200 &  8.080 $\pm$  0.014 &  7.566 $\pm$  0.008 &  7.048 $\pm$  0.014 &  6.035 $\pm$  0.002 & 44.863     & II \\
    04301+2608 & 04331435+2614235 & 68.30956 & 26.23946 & 14.637 & 13.227 & 12.496 & 11.781 $\pm$  0.019 & 11.451 $\pm$  0.024 & 11.059 $\pm$  0.037 &  9.391 $\pm$  0.003 & 71.672     & II \\
        CIDA 14 & 04432023+2940060 & 70.83370 & 29.66794 & 10.402 &  9.730 &  9.408 &  8.894 $\pm$  0.024 &  8.681 $\pm$  0.030 &  8.342 $\pm$  0.003 &  7.622 $\pm$  0.004 & 44.895& II \\
         CX Tau & 04144786+2648110 & 63.69909 & 26.80259 &  9.867 &  9.054 &  8.807 &  8.406 $\pm$  0.015 &  8.103 $\pm$  0.037 &  7.650 $\pm$  0.025 &  6.623 $\pm$  0.018 & 45.922     & II \\
     04216+2603 & 04244457+2610141 & 66.18548 & 26.17016 & 10.796 &  9.753 &  9.053 &  7.780 $\pm$  0.011 &  7.266 $\pm$  0.001 &  6.821 $\pm$  0.019 &  6.000 $\pm$  0.013 & 71.680     & II \\
         UY Aur & 04514737+3047134 & 72.94694 & 30.78664 &  9.134 &  7.987 &  7.239 &  6.046 $\pm$  0.018 & \nodata $\pm$ \nodata &  5.042 $\pm$  0.008 &  3.933 $\pm$  0.003 & 49.359     & II \\
       FV Tau/c & 04265440+2606510 & 66.72671 & 26.11401 & 10.800 &  9.487 &  8.869 &  8.079 $\pm$  0.042 &  7.688 $\pm$  0.031 &  7.150 $\pm$  0.023 &  6.319 $\pm$  0.024 & 71.684     & II \\
         FV Tau & 04265352+2606543 & 66.72303 & 26.11492 &  9.917 &  8.325 &  7.442 &  6.140 $\pm$  0.043 &  5.533 $\pm$  0.024 &  4.932 $\pm$  0.025 &  4.153 $\pm$  0.024 & 71.684     & II \\
         DH Tau & 04294155+2632582 & 67.42285 & 26.54902 &  9.767 &  8.824 &  8.178 &  7.460 $\pm$  0.012 &  7.184 $\pm$  0.012 &  7.146 $\pm$  0.028 &  6.739 $\pm$  0.011 & 71.684     & II \\
         DF Tau & 04270280+2542223 & 66.76133 & 25.70574 &  8.171 &  7.256 &  6.734 &  6.006 $\pm$  0.028 & \nodata $\pm$ \nodata &  5.280 $\pm$  0.008 &  4.638 $\pm$  0.010 & 71.691     & II \\
          IQ Tau & 04295156+2606448 & 67.46455 & 26.11205 &  9.415 &  8.417 &  7.779 &  6.906 $\pm$  0.020 &  6.503 $\pm$  0.008 &  6.191 $\pm$  0.025 &  5.576 $\pm$  0.008 & 71.691     & II \\
       DS Tau A & 04474859+2925112 & 71.95232 & 29.41969 &  9.465 &  8.597 &  8.036 &  7.324 $\pm$  0.041 &  6.909 $\pm$  0.046 &  6.622 $\pm$  0.027 &  6.108 $\pm$  0.039 & 44.898     & II \\
       DS Tau B & 04474810+2925144 & 71.95029 & 29.42064 & 11.889 & 11.545 & 11.453 & 11.425 $\pm$  0.101 & 11.495 $\pm$  0.070 & 11.468 $\pm$  0.070 & \nodata $\pm$  0.084 & 44.898     & III\tablenotemark{b} \\
       DK Tau A & 04304425+2601244 & 67.68439 & 26.02352 &  8.719 &  7.758 &  7.096 &  6.019 $\pm$  0.033 &  5.672 $\pm$  0.024 &  5.410 $\pm$  0.039 &  4.748 $\pm$  0.023 & 71.695     & II \\
       DK Tau B & \nodata & 67.68504 & 26.02320 & \nodata & \nodata & \nodata &  7.815 $\pm$  0.108 &  7.429 $\pm$  0.074 &  7.241 $\pm$  0.111 &  6.608 $\pm$  0.036 & 71.695 & II \\
         SU Aur & 04555938+3034015 & 73.99697 & 30.56673 &  7.199 &  6.558 &  5.990 & \nodata $\pm$ \nodata & \nodata $\pm$ \nodata &  4.638 $\pm$  0.012 &  3.720 $\pm$  0.007 & 49.363     & II \\
         GM Aur & 04551098+3021595 & 73.79527 & 30.36610 &  9.341 &  8.603 &  8.283 &  8.035 $\pm$  0.014 &  7.878 $\pm$  0.006 &  7.700 $\pm$  0.009 &  6.992 $\pm$  0.025 & 49.363     & II \\
       IT Tau A & 04335470+2613275 & 68.47797 & 26.22423 &  9.866 &  8.591 &  7.860 &  7.375 $\pm$  0.091 &  7.111 $\pm$  0.069 &  6.747 $\pm$  0.029 &  6.323 $\pm$  0.036 & 71.656     & II \\
       IT Tau B & \nodata & 68.47744 & 26.22378 & \nodata & \nodata & \nodata &  9.154 $\pm$  0.090 &  8.669 $\pm$  0.061 &  8.197 $\pm$  0.064 &  7.301 $\pm$  0.025 & 71.656     & II \\
         IS Tau & 04333678+2609492 & 68.40302 & 26.16322 & 10.323 &  9.293 &  8.642 &  7.654 $\pm$  0.026 &  7.263 $\pm$  0.024 &  6.859 $\pm$  0.030 &  6.015 $\pm$  0.009 & 71.660     & II \\
       UZ Tau e & 04324303+2552311 & 68.17948 & 25.87520 &  9.136 &  8.117 &  7.354 &  5.978 $\pm$  0.032 &  5.461 $\pm$  0.045 &  5.102 $\pm$  0.029 &  4.334 $\pm$  0.025 & 71.664     & II \\
       UZ Tau w & 04324282+2552314 & 68.17838 & 25.87530 &  9.413 &  8.006 &  7.474 &  7.689 $\pm$  0.045 &  7.552 $\pm$  0.055 &  7.107 $\pm$  0.030 &  6.414 $\pm$  0.024 & 71.664     & II \\
         DL Tau & 04333906+2520382 & 68.41255 & 25.34346 &  9.630 &  8.679 &  7.960 &  6.646 $\pm$  0.033 &  6.124 $\pm$  0.014 &  5.662 $\pm$  0.010 &  4.869 $\pm$  0.006 & 71.664     & II \\
         DO Tau & 04382858+2610494 & 69.61883 & 26.18000 &  9.470 &  8.243 &  7.303 &  6.084 $\pm$  0.053 & \nodata $\pm$ \nodata &  5.070 $\pm$  0.011 &  4.270 $\pm$  0.008 & 71.641     & II \\
         GM Tau & 04382134+2609137 & 69.58858 & 26.15343 & 12.804 & 11.586 & 10.632 &  9.400 $\pm$  0.037 &  8.943 $\pm$  0.025 &  8.629 $\pm$  0.007 &  7.918 $\pm$  0.004 & 71.641     & II \\
    04370+2559 & 04400800+2605253 & 70.03310 & 26.09004 & 12.406 & 10.248 &  8.869 &  7.857 $\pm$  0.010 &  7.292 $\pm$  0.025 &  6.868 $\pm$  0.004 &  5.850 $\pm$  0.004 & 71.645     & II \\
         GN Tau & 04392090+2545021 & 69.83686 & 25.75021 & 10.196 &  8.893 &  8.060 &  7.062 $\pm$  0.011 &  6.656 $\pm$  0.015 &  6.344 $\pm$  0.028 &  5.612 $\pm$  0.004 & 71.648     & II \\
         JH 223 & 04404950+2551191 & 70.20581 & 25.85492 & 10.750 &  9.919 &  9.492 &  8.810 $\pm$  0.017 &  8.511 $\pm$  0.025 &  8.236 $\pm$  0.010 &  7.677 $\pm$  0.006 & 44.863     & II \\
       RW Aur A & 05074953+3024050 & 76.95599 & 30.40105 &  8.378 &  7.621 &  7.020 &  6.268 $\pm$  0.009 &  5.712 $\pm$  0.011 &  5.253 $\pm$  0.018 &  4.345 $\pm$  0.002 & 47.906     & II \\
     V955 Tau & 04420777+2523118 & 70.53225 & 25.38647 &  9.811 &  8.601 &  7.942 &  6.849 $\pm$  0.062 &  6.479 $\pm$  0.032 &  6.056 $\pm$  0.033 &  5.350 $\pm$  0.022 & 44.867     & II \\
     V955 Tau B & 04420732+2523032 & 70.53036 & 25.38409 &  9.580 &  8.401 &  7.945 &  7.640 $\pm$  0.047 &  7.543 $\pm$  0.033 &  7.461 $\pm$  0.024 &  7.439 $\pm$  0.023 & 44.867& III\tablenotemark{b} \\
         CIDA 7 & 04422101+2520343 & 70.58707 & 25.34251 & 11.397 & 10.575 & 10.169 &  9.483 $\pm$  0.007 &  9.105 $\pm$  0.072 &  8.658 $\pm$  0.010 &  7.774 $\pm$  0.001 & 44.871     & II \\
         GO Tau & 04430309+2520187 & 70.76235 & 25.33816 & 10.712 &  9.776 &  9.332 &  8.960 $\pm$  0.033 &  8.622 $\pm$  0.032 &  8.206 $\pm$  0.006 &  7.357 $\pm$  0.008 & 44.871     & II \\
         DP Tau & 04423769+2515374 & 70.65661 & 25.25998 & 10.995 &  9.689 &  8.760 &  7.548 $\pm$  0.020 &  6.909 $\pm$  0.007 &  6.369 $\pm$  0.013 &  5.362 $\pm$  0.003 & 44.875     & II \\
     04414+2506 & 04442713+2512164 & 71.11260 & 25.20420 & 12.195 & 11.359 & 10.761 &  9.717 $\pm$  0.010 &  9.305 $\pm$  0.036 &  8.858 $\pm$  0.009 &  7.861 $\pm$  0.004 & 44.879     & II \\
         CIDA 9 & 05052286+2531312 & 76.34464 & 25.52488 & 12.808 & 11.913 & 11.161 &  9.090 $\pm$  0.038 &  8.555 $\pm$  0.013 &  7.961 $\pm$  0.006 &  6.960 $\pm$  0.004 & 49.375     & II \\
         CIDA 8 & 05044139+2509544 & 76.17193 & 25.16473 & 10.913 & 10.011 &  9.597 &  9.189 $\pm$  0.024 &  8.913 $\pm$  0.028 &  8.665 $\pm$  0.029 &  7.956 $\pm$  0.006 & 49.375     & II \\
        CIDA 11 & 05062332+2432199 & 76.59673 & 24.53845 & 10.421 &  9.712 &  9.459 &  9.002 $\pm$  0.017 &  8.808 $\pm$  0.014 &  8.497 $\pm$  0.009 &  7.810 $\pm$  0.008 & 49.379& II \\
        CIDA 12 & 05075496+2500156 & 76.97855 & 25.00397 & 11.415 & 10.705 & 10.400 & 10.039 $\pm$  0.029 &  9.858 $\pm$  0.006 &  9.525 $\pm$  0.013 &  8.688 $\pm$  0.017 & 49.383     & II \\
         DQ Tau & 04465305+1700001 & 71.72082 & 16.99956 &  9.511 &  8.544 &  7.981 &  7.037 $\pm$  0.002 &  6.569 $\pm$  0.011 &  6.097 $\pm$  0.008 &  5.199 $\pm$  0.004 & 70.852     & II \\
         DR Tau & 04470620+1658428 & 71.77560 & 16.97816 &  8.845 &  7.799 &  6.874 & \nodata $\pm$ \nodata & \nodata $\pm$ \nodata &  4.499 $\pm$  0.008 &  3.786 $\pm$  0.013 & 70.852     & II \\
     Haro6/37  & 04465897+1702381 & 71.74578 & 17.04369 &  9.239 &  7.991 &  7.310 &  6.581 $\pm$  0.073 &  6.170 $\pm$  0.038 &  5.802 $\pm$  0.022 &  5.228 $\pm$  0.023 & 70.852     & II \\
      Haro6/37c & \nodata  & 71.74628 & 17.04428 & \nodata & \nodata & \nodata &  7.838 $\pm$  0.078 &  7.320 $\pm$  0.073 &  7.041 $\pm$  0.033 &  5.988 $\pm$  0.024 & 70.852     & II \\
         FP Tau & 04144730+2646264 & 63.69677 & 26.77361 &  9.897 &  9.175 &  8.873 &  8.455 $\pm$  0.011 &  8.194 $\pm$  0.017 &  7.880 $\pm$  0.008 &  7.218 $\pm$  0.031 & 45.922& II \\
        LkCa 14 & 04361909+2542589 & 69.07929 & 25.71599 &  9.336 &  8.713 &  8.580 &  8.474 $\pm$  0.034 &  8.540 $\pm$  0.018 &  8.485 $\pm$  0.005 &  8.441 $\pm$  0.005 & 71.652& III \\
     CoKu Tau/4 & 04411681+2840000 & 70.31945 & 28.66627 & 10.163 &  9.077 &  8.656 &  8.446 $\pm$  0.010 &  8.394 $\pm$  0.020 &  8.321 $\pm$  0.007 &  8.116 $\pm$  0.003 & 44.895    & III \\
       FW Tau A & 04292971+2616532 & 67.37374 & 26.28136 & 10.340 &  9.675 &  9.388 &  8.957 $\pm$  0.053 &  8.964 $\pm$  0.045 &  8.881 $\pm$  0.026 &  8.819 $\pm$  0.024 & 71.688    & III\tablenotemark{b} \\
       FW Tau B & 04292887+2616483 & 67.37028 & 26.28007 & 10.784 &  9.792 &  9.417 &  9.204 $\pm$  0.049 &  9.200 $\pm$  0.042 &  9.104 $\pm$  0.030 &  9.030 $\pm$  0.023 & 71.688    & III\tablenotemark{b} \\
         DI Tau & 04294247+2632493 & 67.42670 & 26.54652 &  9.323 &  8.599 &  8.391 &  8.186 $\pm$  0.033 &  8.195 $\pm$  0.026 &  8.179 $\pm$  0.014 &  8.091 $\pm$  0.008 & 71.684    & III \\
       LkH332G1 & 04420732+2523032 & 70.53003 & 25.38376 &  9.580 &  8.401 &  7.945 &  7.602 $\pm$  0.024 &  7.534 $\pm$  0.020 &  7.509 $\pm$  0.052 &  7.399 $\pm$  0.027 & 44.867& III \\
      LkH332/G2 & 04420548+2522562 & 70.52239 & 25.38183 &  9.787 &  8.663 &  8.227 &  7.877 $\pm$  0.015 &  7.834 $\pm$  0.022 &  7.751 $\pm$  0.007 &  7.660 $\pm$  0.005 & 44.867& III \\
        LkCa 19 & 04553695+3017553 & 73.90353 & 30.29824 &  8.870 &  8.318 &  8.148 &  8.078 $\pm$  0.017 &  8.128 $\pm$  0.020 &  8.087 $\pm$  0.010 &  8.025 $\pm$  0.005 & 49.367& III \\
        CIDA 13 & 04391586+3032074 & 69.81579 & 30.53504 & 12.680 & 12.069 & 11.834 & 11.554 $\pm$  0.026 & 11.538 $\pm$  0.007 & 11.455 $\pm$  0.058 & \nodata $\pm$ \nodata & 73.113& III \\
  RXJ04467+2459 & 04464260+2459034 & 71.67705 & 24.98389 & 11.261 & 10.667 & 10.338 &  9.980 $\pm$  0.006 &  9.929 $\pm$  0.001 &  9.896 $\pm$  0.012 &  9.837 $\pm$  0.012 & 44.883& III \\
        HBC 360 & 04043936+2158186 & 61.16364 & 21.97144 & 10.798 & 10.171 &  9.966 &  9.624 $\pm$  0.052 &  9.589 $\pm$  0.019 &  9.517 $\pm$  0.029 &  9.473 $\pm$  0.022 & 46.426    & III \\
        HBC 361 & 04043984+2158215 & 61.16562 & 21.97226 & 10.939 & 10.353 & 10.101 &  9.728 $\pm$  0.009 &  9.704 $\pm$  0.015 &  9.624 $\pm$  0.026 &  9.572 $\pm$  0.020 & 46.426    & III \\
       J 4872 B & 04251767+2617504 & 66.32294 & 26.29675 &  9.537 &  8.475 &  8.545 &  9.007 $\pm$  0.062 &  8.942 $\pm$  0.027 &  8.877 $\pm$  0.030 &  8.848 $\pm$  0.022 & 71.699    & III \\
       J 4872 A & \nodata  & 66.32377 & 26.29731 & \nodata & \nodata & \nodata &  8.307 $\pm$  0.042 &  8.337 $\pm$  0.044 &  8.230 $\pm$  0.022 &  8.164 $\pm$  0.023 & 71.699    & III \\
          JH 56 & 04311444+2710179 & 67.80986 & 27.17122 &  9.705 &  9.036 &  8.794 &  8.703 $\pm$  0.027 &  8.742 $\pm$  0.017 &  8.707 $\pm$  0.010 &  8.631 $\pm$  0.007 & 71.715    & III \\
          J 507 & 04292071+2633406 & 67.33598 & 26.56081 &  9.821 &  9.089 &  8.791 &  8.521 $\pm$  0.014 &  8.497 $\pm$  0.028 &  8.446 $\pm$  0.005 &  8.398 $\pm$  0.007 & 71.699    & III \\
        HBC 427 & 04560201+3021037 & 74.00796 & 30.35061 &  8.958 &  8.317 &  8.129 &  8.056 $\pm$  0.018 &  8.095 $\pm$  0.025 &  8.055 $\pm$  0.013 &  8.001 $\pm$  0.004 & 49.371    & III \\
          J 665 & 04315844+2543299 & 67.99329 & 25.72457 & 10.585 &  9.828 &  9.559 &  9.283 $\pm$  0.019 &  9.243 $\pm$  0.070 &  9.257 $\pm$  0.015 &  9.200 $\pm$  0.006 & 71.668    & III \\
       HV Tau AB & 04383528+2610386 & 69.64704 & 26.17735 & 9.227 & 8.284 &  7.906 &  7.644 $\pm$  0.094 &  7.578 $\pm$  0.036 &  7.468 $\pm$  0.022 &  7.441 $\pm$  0.025 & 71.641    & III \\
       HV Tau C & \nodata  & 69.64792 & 26.17816 & \nodata & \nodata & \nodata & 11.327 $\pm$  0.139 & 10.743 $\pm$  0.045 & 10.223 $\pm$  0.045 &  9.379 $\pm$  0.040 & 71.641    & II\tablenotemark{b} \\
         IW Tau & 04410470+2451062 & 70.26911 & 24.85132 &  9.244 &  8.479 &  8.275 &  8.067 $\pm$  0.018 &  8.092 $\pm$  0.012 &  8.022 $\pm$  0.002 &  7.972 $\pm$  0.003 & 44.879    & III \\
         J2 157 & 04205273+1746415 & 65.21945 & 17.77789 & 11.618 & 11.040 & 10.776 & 10.401 $\pm$  0.017 & 10.383 $\pm$  0.017 & 10.356 $\pm$  0.008 & 10.278 $\pm$  0.003 & 72.336    & III \\
        HBC 372 & 04182147+1658470 & 64.58923 & 16.97939 & 11.175 & 10.603 & 10.464 & 10.355 $\pm$  0.003 & 10.443 $\pm$  0.015 & 10.405 $\pm$  0.036 & 10.313 $\pm$  0.022 & 72.340    & III \\
        CIDA 10 & 05061674+2446102 & 76.56924 & 24.76910 & 10.792 & 10.091 &  9.815 &  9.522 $\pm$  0.018 &  9.508 $\pm$  0.028 &  9.457 $\pm$  0.018 &  9.393 $\pm$  0.027 & 49.383& III \\
  RXJ05072+2437 & 05071206+2437163 & 76.79977 & 24.62079 & 10.141 &  9.490 &  9.297 &  9.220 $\pm$  0.023 &  9.209 $\pm$  0.050 &  9.208 $\pm$  0.006 &  9.144 $\pm$  0.006 & 49.387& III \\
          KPNO5 & 04294568+2630468 & 67.43999 & 26.51254 & 12.640 & 11.918 & 11.536 & 11.036 $\pm$  0.015 & 10.951 $\pm$  0.006 & 10.868 $\pm$  0.031 & 10.826 $\pm$  0.028 & 71.684  &  III\tablenotemark{b} \\
          KPNO3 & 04262939+2624137 & 66.62216 & 26.40343 & 13.323 & 12.501 & 12.079 & 11.319 $\pm$  0.031 & 10.978 $\pm$  0.014 & 10.535 $\pm$  0.010 &  9.618 $\pm$  0.003 & 71.703  &  II\tablenotemark{b} \\
          KPNO4 & 04272799+2612052 & 66.86636 & 26.20104 & 14.997 & 14.025 & 13.281 & 12.531 $\pm$  0.015 & 12.381 $\pm$  0.024 & 12.270 $\pm$  0.160 & 12.069 $\pm$  0.319 & 71.703  &  III\tablenotemark{b} \\
          KPNO7 & 04305718+2556394 & 67.73805 & 25.94391 & 14.521 & 13.828 & 13.272 & 12.577 $\pm$  0.017 & 12.311 $\pm$  0.039 & 11.979 $\pm$  0.071 & 11.194 $\pm$  0.062 & 71.695  &  II\tablenotemark{b} \\
          KPNO6 & 04300724+2608207 & 67.52984 & 26.13874 & 14.995 & 14.197 & 13.689 & 13.033 $\pm$  0.033 & 12.781 $\pm$  0.036 & 12.535 $\pm$  0.168 & 11.770 $\pm$  0.114 & 71.711  &  II\tablenotemark{b} \\
         TW Hya & 11015191-3442170 &165.46632 &-34.70472 & 8.217 & 7.558 & 7.297  &  7.13  $\pm$   0.03 &  7.07  $\pm$  0.01  &  6.93  $\pm$ 0.02  &   5.98 $\pm$ 0.01 & \nodata\tablenotemark{c} & II \\
\enddata 
\tablenotetext{a}{Used 2MASS extended source magnitudes for DG Tau B}
\tablenotetext{b}{New classification, based on IRAC data in this paper; $^{\rm c}$Observation date JD $=$ 52993.131}
\tablecomments{The position of CIDA 12 in Brice\~no \etal (1993) is in error;
should be (J2000) 5h 07m 54.98s +25d 00m 15s.}
\end{deluxetable}

\end{document}